\documentclass[aps, prd, twocolumn,showpacs, superscriptaddress, groupedaddress]{revtex4} 
\usepackage{graphicx}
\usepackage{gensymb}
\usepackage{amssymb}
\usepackage{amsmath}
\usepackage{dcolumn}
\usepackage{color}
\usepackage[utf8]{inputenc}
\usepackage{subfigure, rotating, bm, array}
\usepackage[pagebackref=false, colorlinks=true]{hyperref}
\hypersetup{
linkcolor=blue,     
citecolor=blue,     
urlcolor=blue}      
\begin{document}
\title{Lense-Thirring effect and precession of timelike geodesics in 
slowly rotating black hole and naked singularity spacetimes}

\author{Parth Bambhaniya}
\email{grcollapse@gmail.com}
\affiliation{International Center for Cosmology, Charusat University, Anand, GUJ 388421, India}
\author{Jay S. Verma}
\email{jay.verma2210@gmail.com}
\affiliation{International Center for Cosmology, Charusat University, Anand, GUJ 388421, India}
\author{Dipanjan Dey}
\email{dipanjandey.icc@charusat.ac.in}
\affiliation{International Center for Cosmology, Charusat University, Anand, GUJ 388421, India}
\author{Pankaj S. Joshi}
\email{psjprovost@charusat.ac.in}
\affiliation{International Center for Cosmology, Charusat University, Anand, GUJ 388421, India}
\author{Ashok B. Joshi}
\email{gen.rel.joshi@gmail.com}
\affiliation{International Center for Cosmology, Charusat University, Anand, GUJ 388421,  India}
\affiliation{}

\date{\today}

\begin{abstract}
In this paper, we investigate the properties of the Lense-Thirring precession frequency and periastron precession of the relativistic bound orbits in slowly rotating naked singularity spacetimes. We show that the precession frequencies of a stationary test gyroscope are quite unique in slowly rotating naked singularity spacetimes, and therefore, one can easily distinguish them from the slowly rotating Kerr black hole spacetime.   On the other hand, we also show that  in slowly rotating Joshi-Malafarina-Narayan (JMN1) naked singularity spacetime, negative precession of the timelike bound orbits is allowed which is not possible in Kerr and slowly rotating null naked singularity spacetimes.


\end{abstract}

\maketitle

\section{Introduction}
It is well known that the end state of the continual gravitational collapse of matter cloud is the spacetime singularity \cite{hawking, penrose}. However, there is no such theorem on the causal nature of spacetime singularity formed during the gravitational collapse. In 1969, Penrose gave the cosmic censorship conjecture (CCC) which does not allow a spacetime singularity to be causally connected with other spacetime points \cite{penrose2}. There are two versions of the cosmic censorship conjecture: weak cosmic censorship conjecture and strong cosmic censorship conjecture. The weak cosmic censorship conjecture does not allow any globally visible singularity \cite{tiplerclarkeellis}, whereas its strong counter-part states that a spacetime singularity should not be locally visible as well \cite{geroch, hawking2, penrose3}. One can show that a globally hyperbolic spacetime is necessary for the validity of strong cosmic censorship conjecture \cite{psjoshi2,joshinarlikar, bruhat, klainerman, ringstrom, sbierski}. It can be shown that the singularity formed during the homogeneous dust (i.e. pressure is zero) collapse is always locally hidden. However, if one introduces some non-zero pressure or inhomogeneity in density, then with some suitable initial data, it can be shown that the singularity can be locally as well as globally visible. In many literature, the formation of the visible or naked singularity during the gravitational collapse of matter cloud is shown \cite{eardley, christodoulou, joshi, goswami, joshi2, vaz, jhingan0, deshingkar, mena, magli1, magli2, giambo1, harada1, harada2, joshi7, mosani1, mosani2, mosani3, mosani4, mosani5}. In \cite{Joshi:2011zm}, authors show that a spacetime with a central naked singularity can be formed as an equilibrium end state of the gravitational collapse of general matter cloud. Therefore, in the regime of classical general relativity, there are many theoretical scenarios where a visible singularity exists. Now, the important question is whether they really exist in nature and if they really exist then what would be their physical signature.

In several papers, the possible physical signatures of the ultra-high density compact objects are discussed in detail \cite{Shaikh:2019hbm,Gralla:2019xty,Abdikamalov:2019ztb,Yan:2019etp,Vagnozzi:2019apd,Gyulchev:2019tvk,Shaikh:2019fpu,Dey:2013yga,Dey+15,Shaikh:2018lcc,Joshi2020,Paul2020,Dey:2020haf,Dey:2020bgo,abdujabbarov_2015a, atamurotov_2015, abdujabbarov_2017, abdujabbarov_2015b, younsi_2016, papnoi_2014, bambi_2013a, ohgami_2015, stuchlik_2018, stuchlik_2019,Martinez:2019nor, Eva,Eva1,Eva2,tsirulev,Joshi:2019rdo,Bhattacharya:2017chr,Bambhaniya:2019pbr,Dey:2019fpv,Bam2020,Bambhaniya:2021ybs,Lin:2021noq,Deng:2020yfm,Deng:2020hxw,Gao:2020wjz,aa4}. In \cite{Shaikh:2019hbm,Gralla:2019xty,Abdikamalov:2019ztb,Yan:2019etp,Vagnozzi:2019apd,Gyulchev:2019tvk,Shaikh:2019fpu,Dey:2013yga,Dey+15,Shaikh:2018lcc,Joshi2020,Paul2020,Dey:2020haf,Dey:2020bgo,abdujabbarov_2015a, atamurotov_2015, abdujabbarov_2017, abdujabbarov_2015b, younsi_2016, papnoi_2014, bambi_2013a, ohgami_2015, stuchlik_2018, stuchlik_2019}, authors elaborately discuss the shadow properties of different types of compact objects. With respect to an asymptotic observer, the image of a compact object may consist of a central dark region surrounded by a high intensity or bright region, where the intensity maximizes at the boundary of the dark region and smoothly decreases as radial distance increases. In the asymptotic observer's sky, the central dark region is the shadow of the compact object. In several recent papers \cite{Shaikh:2018lcc,Joshi2020,Paul2020,Dey:2020haf,Dey:2020bgo}, it is shown that the shadow is not only the property of the black hole but it can also be cast by other horizon less compact objects. In \cite{Shaikh:2018lcc}, authors show that a naked singularity spacetime can cast shadow of similar size to what an equally massive Schwarzschild black hole can cast. In \cite{Joshi2020, Dey:2020bgo}, the shadow properties of the nulllike and timelike naked singularities are elaborately discussed. In \cite{Dey:2020haf}, a spacetime configuration having a central naked singularity is considered and it is shown that a thin shell of matter can cast shadow in the absence of photon sphere as well as event horizon. All these theoretical studies on the shadow of ultra-compact objects are very much important in the context of recent observation of the shadow of M87 glactic center and the upcoming image of Milky-way galactic center Sgr-A*.

Another important physical signature of a horizon-less ultra-compact object is the orbital dynamics of stars around it which can distinguish it from the black holes. As we know, the nature of timelike geodesics in a spacetime depends upon the geometry of that spacetime.
Therefore, the trajectory of stars around a compact object can reveal the causal structure of that compact object. Hence, the stellar dynamics close to our Milky-way galactic center can reveal important properties of Sgr-A*. For this purpose, GRAVITY, SINFONI, UCLA galactic group are continuously tracking the dynamics of `S' stars which are orbiting around Sgr-A* with very small (w.r.t galactic scale) pericenter distances \cite{M87, Eisenhauer:2005cv, center1}. In this context, there are many literature where the timelike geodesics around different types of compact objects are elaborately discussed \cite{Martinez:2019nor, Eva,Eva1,Eva2,tsirulev,Joshi:2019rdo,Bhattacharya:2017chr,Bambhaniya:2019pbr,Dey:2019fpv,Bam2020,Bambhaniya:2021ybs,Lin:2021noq,Deng:2020yfm,Deng:2020hxw,Gao:2020wjz,aa4}. In some of the recent papers, we show that the timelike bound orbits of a massive particle in the presence of a central naked singularity, can precess in the opposite direction of the particle motion and we denote that type of precession as negative precession. It is shown that in Schwarzschild and Kerr black hole spacetime, the negative precession of the timelike bound orbits is not allowed. However, there are some naked singularity spacetimes for which both the positive and negative precession of bound orbits are possible \cite{Bambhaniya:2019pbr,Dey:2019fpv,Bam2020,Bambhaniya:2021ybs}. Therefore, the nature of the precession of the bound timelike orbits can give us the information about the causal structure of the central compact object. Accretion disk around a compact object and its thermal property is also an important physical phenomenon that can be used for distinguishing black holes from other compact objects \cite{Guo:2020tgv}.

In this paper, we discuss the nature of Lense-Thirring precession of a test gyro and the orbital precession of a test particle in two rotating naked singularity spacetimes, where we consider slow rotation approximation. We use the Newman-Janis algorithm \cite{Janis Newman, Drake:1998gf} to construct the rotating counterparts of the static 1st type of Joshi-Malafarina-Narayan (JMN1) spacetime \cite{Joshi:2011zm} and a null naked singularity spacetime \cite{Joshi2020}. We investigate the frame-dragging effect of these two rotating naked singularity spacetimes on the spin of a test gyro (i.e. the Lense-Thirring precession ). The Lense-Thirring precession is one of the important prediction of Einstein's general relativity and in the past, a huge amount of efforts were given to verify the prediction \cite{lt, schiff1}. As it is known, on 4th May 2011, NASA and the Standford-based analysis group confirmed that the Gravity Probe B satellite measured an equal amount of precession that can be predicted considering the general relativistic frame dragging effect of Earth \cite{ev}. In Gravity Probe B, four very sensitive gyroscopes are installed and initially, they were pointed towards a fixed star HR8703 which is also known as IM Pegasi. However, due to the earth's rotation, the frame dragging effect changes the direction of the gyroscopes. Due to the weak gravitational field of the earth (compare to ultra-high compact object), the Gravity Probe B recorded a precession rate $37.2\pm 7.2$ milliarcsecond/ year, whereas general relativistic prediction is $39.2\pm 0.19$ milliarcsecond/ year. 
After the confirmation of Gravity Probe B, the precession of a test gyroscope near the rotating ultra-high-density compact objects has become an important field of research, where a pulsar near a compact object plays the role of the gyroscope \cite{Chakraborty:2016ipk,Chakraborty:2016mhx,Kocherlakota:2017hkn}. Due to the Lense-Thirring effect of a rotating massive body, the pulse rate of a pulsar is expected to be varied and therefore, the observational results on the pulse rate of a pulsar near an ultra-compact object can reveal the causal structure of that compact object \cite{Kocherlakota:2017hkn}.
Along with the Lense-Thirring precession, in this paper, we also discuss the orbital precession of a test particle around the rotating JMN1 and null naked singularity spacetime and we show that the rotating counterpart of JMN1 spacetime also allows negative precession of bound orbits which is not possible in Kerr spacetime.

This paper is organised as follows. In Sec.~(\ref{JN1}), we discuss the Newman-Janis algorithm and show how one can derive the rotating version of a spherically symmetric and static spacetime using that algorithm. In that section, we also discuss why we need to consider the slow rotation approximation to get the rotating form of a general static, spherically symmetric spacetime using Newman-Janis algorithm. At last, in that section, we construct the slowly rotating forms of JMN1 and null naked singularity spacetimes. In Sec.~(\ref{LTeffect}), we investigate the Lense-Thirring effects on a test gyroscope in those two rotating naked singularity spacetimes and compare the results with that for Kerr spacetime. In Sec.~(\ref{orbit1}), we discuss the nature of the precession of the timelike bound orbits in slowly rotating JMN1 and the null naked singularity spacetimes. In Sec.~(\ref{concludesec}), we conclude by discussing the important results of this paper. Throughout this paper, we consider Newton's gravitational constant $G_N$ and light velocity $c$ as unity.
 
\section{The general Newman-Janis procedure}
\label{JN1}
In \cite{Janis Newman}, Janis and Newman develop a procedure by which the Kerr metric can be obtained from the Schwarzschild metric. In this section, at first, we briefly discuss the general Newman-Janis procedure (NJP) for the readers' convenience and then we construct the rotating versions of JMN1 and null naked singularity spacetime using that algorithm. 
The NJP consists of a four-step method which help us to construct stationary, axially symmetric, and rotating spacetime from a seed metric. In this paper, we consider the seed metric as static and spherically symmetric. 
The general spherically symmetric and static metric in Boyer-Lindquist coordinates (BLC) can be written as,
\begin{eqnarray} \label{eq:Seed_Metrics}
\text{d}s^2 = -f(r)~\text{d}t^2 + \frac{~\text{d}r^2}{g(r)}  + h(r)\text{d}\Omega^2,
\end{eqnarray}
where $\text{d}\Omega^2 = d\theta^2+\sin^2\theta d\phi^2$. Now, the first step is to write the seed metric in outgoing or incoming Eddington-Finkelstein coordinates ($u,r,\theta,\phi$). Following is the transformation from BLC to outgoing Eddington-Finkelstein coordinate system,
\begin{equation}
du=dt-\frac{dr}{\sqrt{f(r)g(r)}}.
\end{equation}
Therefore, in these advanced Eddington-Finkelstein coordinates, the metric (\ref{eq:Seed_Metrics}) is expressed as,
\begin{equation}
\text{d}s^2 = -f(r)~\text{d}u^2-2\sqrt{\frac{f(r)}{g(r)}}~\text{d}u~\text{d}r + h(r)\text{d}\Omega^2.
\end{equation}
In step two, we introduce the inverse metric $g^{\mu\nu}$ using the null tetrad $(l^\mu,n^\mu,m^\mu,\bar{m}^\mu)$ as,
\begin{equation}
g^{\mu\nu}=-l^\mu n^\nu -l^\nu n^\mu +m^\mu \bar{m}^\nu +m^\nu \bar{m}^\mu,
\end{equation}
where over-line indicates complex conjugation and these null tetrads can be written as,
\begin{eqnarray}
l^\mu=\delta^\mu_r, \hspace{0.2cm} n^\mu&=&\sqrt{\frac{g(r)}{f(r)}}\delta^\mu_u-\frac{g(r)}{2}\delta^\mu_r,\nonumber\\  m^\mu &=&\frac{1}{\sqrt{2}r}\left(\delta^\mu_\theta+\frac{i}{\sin\theta}\delta^\mu_\phi\right),\nonumber\\
\bar{m}^\mu &=&\frac{1}{\sqrt{2}r}\left(\delta^\mu_\theta-\frac{i}{\sin\theta}\delta^\mu_\phi\right),
\end{eqnarray}
where they satisfy the following relations,
\begin{equation}
l_\mu l^\mu = n_\mu n^\mu = m_\mu m^\mu = l_\mu m^\mu = n_\mu m^\mu =0,
\end{equation}
\begin{equation}
l_\mu n^\mu = - m_\mu \bar{m}^\mu =-1.
\end{equation}
In order to obtain a new axially symmetric and rotating spacetime, one need to do complexification of the functions as, $f(r), g(r), h(r)$ $\rightarrow \tilde{f}(r^\prime), \tilde{g}(r^\prime), \tilde{h}(r^\prime)$. Therefore, the third step is the complexification of coordinates in which we allow the coordinates to take complex values. The radial coordinate r can be written as a real function of r and its complex conjugate $\bar{r}$. Following are the examples of that modifications,
\begin{equation}
\frac{1}{r}\rightarrow \frac{1}{2}\left(\frac{1}{r'}+\frac{1}{\bar{r}'}\right), \hspace{0.3cm} r^2\rightarrow r'\bar{r}',
\label{rmod}
\end{equation}
where the complex coordinate can be written as, 
\begin{eqnarray}
u^{\prime}=u-i a \cos\theta,\,\, r^{\prime}=r+i a \cos\theta,\,\, \theta^{\prime}=\theta,\,\, \phi^{\prime}=\phi\, .
\end{eqnarray}
The reason to choose this particular complexification (Eq.~(\ref{rmod})) is that it is successful in generating the rotating vacuum solutions. 
In the \cite{Drake:1998gf}, it is shown that this is the unique choice of complexification to generate Kerr-Newman solution from Reissner–Nordstrom spacetime using the Newman-Janis procedure. 
With the complex coordinate system, the null tetrads can be written as,
\begin{eqnarray}
 l'^\mu &=& \delta^\mu_r,\hspace{0.3cm}  n'^\mu=\sqrt{\frac{\widetilde{g}(r')}{\widetilde{f}(r')}}\delta^\mu_u-\frac{\widetilde{g}(r')}{2}\delta^\mu_r,\nonumber \\
 m^\mu &=&\frac{1}{\sqrt{2\widetilde{h}(r')}}\left(ia\sin\theta(\delta^\mu_u-\delta^\mu_r)+\delta^\mu_\theta+\frac{i}{\sin\theta}\delta^\mu_\phi\right),
 \nonumber\\
 \bar{m^\mu} &=& \frac{1}{\sqrt{2\widetilde{h}(r')}}\left(ia\sin\theta(\delta^\mu_u-\delta^\mu_r)-\delta^\mu_\theta+\frac{i}{\sin\theta}\delta^\mu_\phi\right).\nonumber\\
\end{eqnarray}
Now, using the new null tetrads, one can write the metric in the advance Eddington-Finkelstein or null coordinates as,
\begin{widetext}
\begin{eqnarray}
 \label{eq:null_coordinate_metric_1}
ds^2 =&\ -\widetilde{f}(r)du^2-2\sqrt{\frac{\widetilde{f}(r)}{\widetilde{g}(r)}}dudr+2a\sin^2\theta\left(\widetilde{f}(r)-\sqrt{\frac{\widetilde{f}(r)}{\widetilde{g}(r)}}\right)dud\phi  +2a\sqrt{\frac{\widetilde{f}(r)}{\widetilde{g}(r)}}\sin^2\theta drd\phi \nonumber\\ 
&\ +\widetilde{h}(r)d\theta^2 
+\sin^2\theta\left[\widetilde{h}(r)+a^2\sin^2\theta\left(2\sqrt{\frac{\widetilde{f}(r)}{\widetilde{g}(r)}}-\widetilde{f}(r)\right)\right]d\phi^2.
\end{eqnarray} 
\end{widetext}
Here, one can see, the complexification usually produces a metric with several non-zero off-diagonal components.
These can be removed (except $dt d\phi$) by transforming it in the Boyer-Lindquist coordinates $(t,r,\theta,\phi)$. 
So, the final step of the Newman-Janis algorithm is to write down the metric in Boyer-Lindquist form using the following transformations,
\begin{equation}
du=dt'+F(r)dr, \hspace{0.5cm} d\phi=d\phi'+G(r) dr.
\label{eq:transformation_to_BL}
\end{equation}
Inserting the above transformations in the metric (\ref{eq:null_coordinate_metric_1}) and setting $g_{tr}$ and $g_{r\phi}$ to zero, we obtain
\begin{equation}
F(r)=-\frac{\sqrt{\frac{\widetilde{g}(r,\theta)}{\widetilde{f}(r,\theta)}}\widetilde{h}(r,\theta)+a^2\sin^2\theta}{\widetilde{g}(r,\theta)\widetilde{h}(r,\theta)+a^2\sin^2\theta},
\label{eq:F}
\end{equation}
\begin{equation}
G(r)=-\frac{a}{\widetilde{g}(r,\theta)\widetilde{h}(r,\theta)+a^2\sin^2\theta}.
\label{eq:G}
\end{equation}
\begin{figure*}
\centering
\subfigure[Figure for $ M_{0} = 0.10907099 $ and $R_{b} = 18.336 $.]
{\includegraphics[width=65mm]{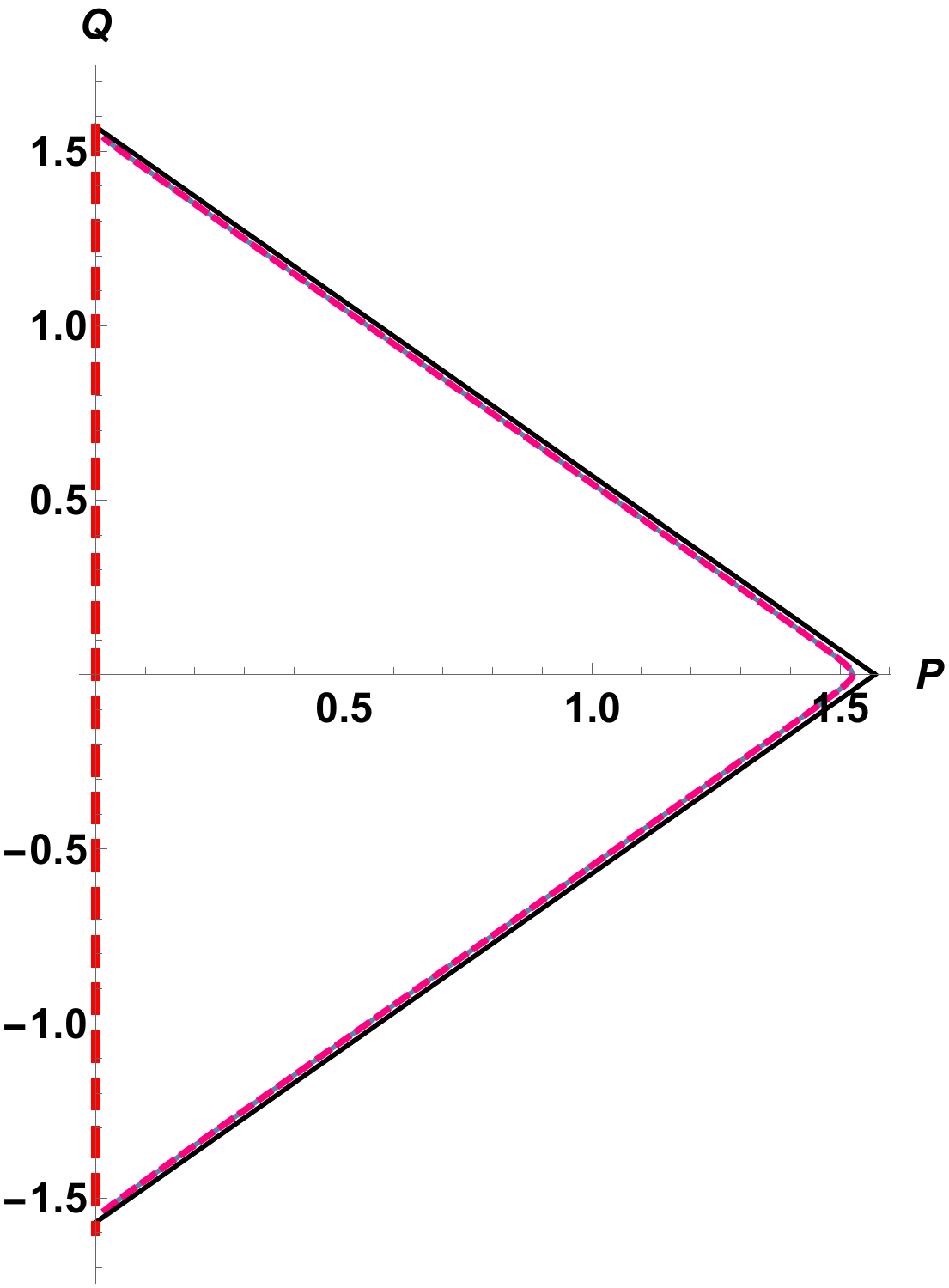}\label{re3}}
\subfigure[Figure for $ M_{0} = 0.9579581 $ and $R_{b} = 2.08778 $.]
{\includegraphics[width=87mm]{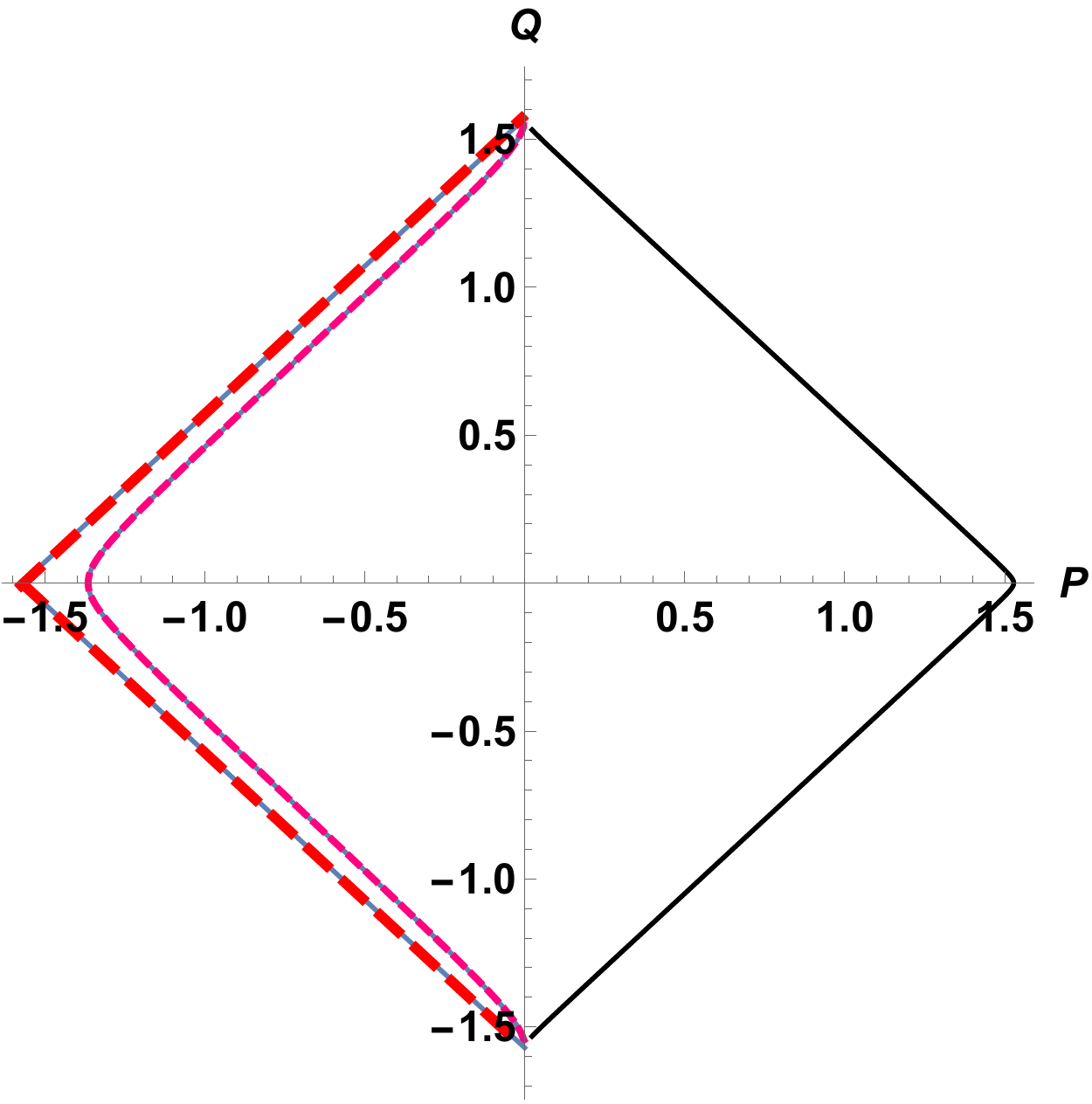}\label{re4}}\\
 \caption{In this figure, we show the Penrose diagrams of JMN1 null and timelike naked singularities. In Figs. (\ref{re3}) and (\ref{re4}), we show the Penrose diagrams of JMN1 spacetime with $ M_{0}<2/3$ and $ M_{0}>2/3$ respectively. It can be seen that for $M_0 > 2/3$, the central singularity is null like and for $M_0 < 2/3$, the singularity becomes timelike (i.e. shown by the red dashed line). In the above figures, the matching radius ( $R_b$ ) of a junction between internal JMN1 and external Schwarzschild spacetime is shown by the pink dashed line, whereas, the black line represents asymptotic regions in Schwarzschild spacetime. Therefore, the area in between the pink dashed line and the black line represents Schwarzschild spacetime, and the area in between the red dashed line and the pink dashed line represents JMN1 spacetime.}\label{fig1}
\end{figure*}
Note that the transformation in (\ref{eq:transformation_to_BL}) is possible only when the left hand sides of Eqs. (\ref{eq:F}) and (\ref{eq:G}) are independent of $\theta$. We find the transformation in (\ref{eq:transformation_to_BL}) is not possible in general. However, to construct a valid transformation, one needs to consider the slow rotation limit $(a^2<<r^2)$, where we ignore the $a^2$ and higher order terms of $a$. Therefore, in the slow rotation limit, the metric functions $f(r), g(r)$ and $h(r)$ remain independent of $\theta$ after complexification. Additionally, to make the right hand side of Eqs. (\ref{eq:F}) and (\ref{eq:G}) independent of $\theta$, we have to assume the other conditions for the slow rotation limit as below,
\begin{eqnarray}
 a^2\ll \sqrt{\frac{g(r)}{f(r)}}h(r), \hspace{0.3cm}
 a^2\ll g(r)h(r).
\end{eqnarray}
In the slow rotation limit, by substituting the transformation (\ref{eq:transformation_to_BL}) in (\ref{eq:null_coordinate_metric_1}), we can write the general form of the slow rotating spacetime as,
\begin{eqnarray}\label{general_form}
 ds^2 = -f(r)dt^2&+&\frac{dr^2}{g(r)}+r^2\left(d\theta^2+\sin^2\theta d\phi^2\right)\nonumber\\
 &-&2a\sin^2\theta\left(\sqrt{\frac{f(r)}{g(r)}}-f(r)\right) dtd\phi.\nonumber\\
\end{eqnarray}
\begin{figure*}
    \centering
    \includegraphics[scale=0.65]{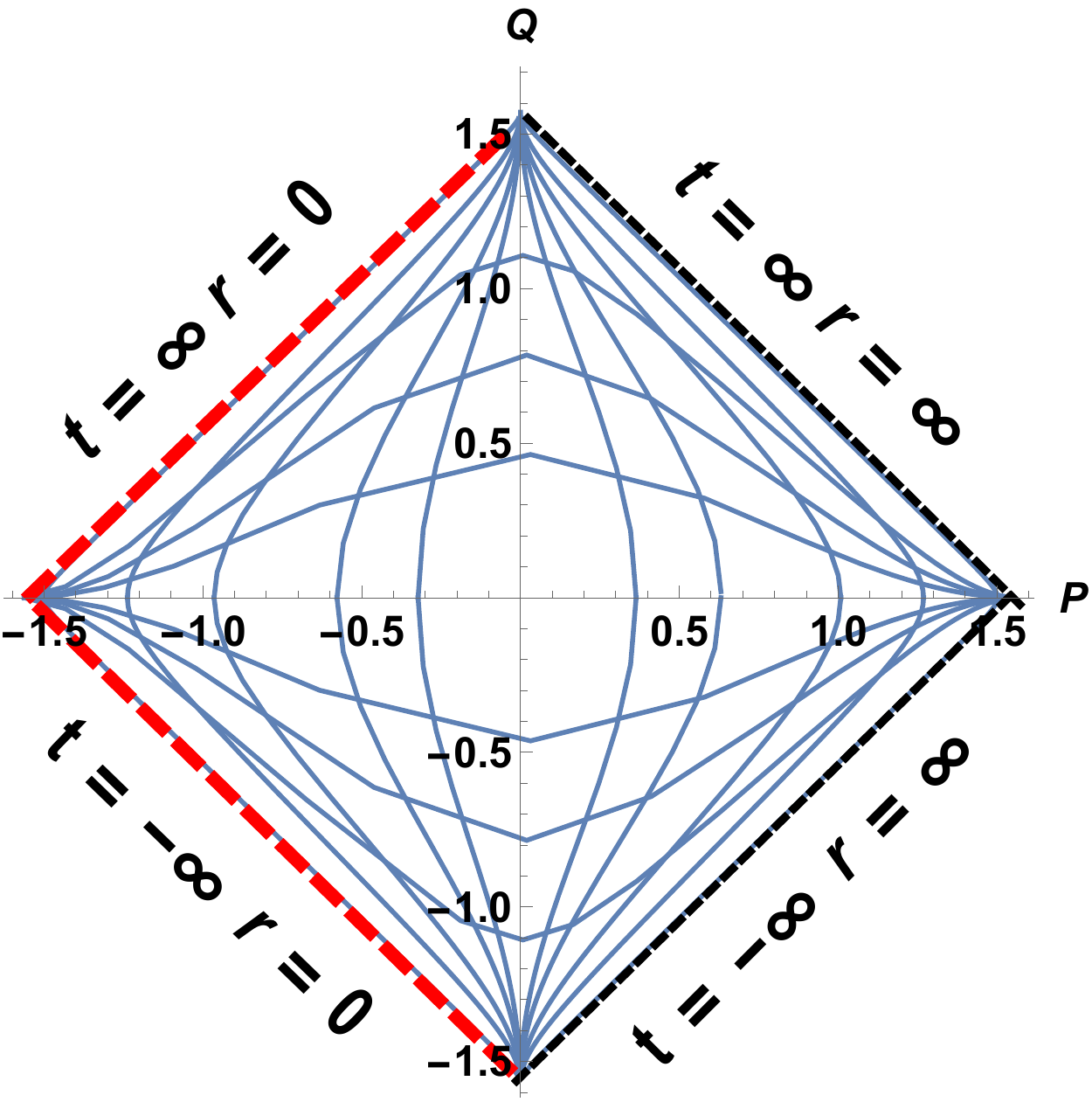}
    \caption{Penrose diagram of null naked singularity}
    \label{fig2}
\end{figure*}

\subsection{Joshi-Malafarina-Narayan 1 (JMN1) Naked Singularity Spacetime}
\label{JMNspt}
As we know, the Oppenheimer-Snyder and Datt (OSD) collapse model describes the continual gravitational collapse of a spherically symmetric, homogeneous dust cloud. The end state of the OSD collapse is always a black hole. However, as it was discussed previously, for the matter cloud having inhomogeneous density and pressures, the final fate of the gravitational collapse can be either a naked singularity or a black hole depending upon the initial conditions. Apart from the above mentioned scenarios, such scenarios can arise where the presence of pressure in matter cloud can equilibrate the collapsing body into a static configuration in a large co-moving time. In \cite{Joshi:2011zm}, it is shown that a static naked singularity spacetime, namely 1st type of Joshi-Malafarina-Narayan (JMN1) spacetime, can be formed as an end equilibrium state of gravitational collapse. JMN1 spacetime is a spherically symmetric static spacetime which has a central naked singularity. This spacetime is seeded by anisotropic matter which has zero radial and non-zero tangential pressures. The line element of this metric can be written as, 

\begin{eqnarray} 
 ds^2 = -(1- M_0) \left(\frac{r}{R_b}\right)^\frac{M_0}{(1- M_0)}dt^2 + \frac{dr^2}{(1 - M_0)} + r^2d\Omega^2\,\, , \nonumber\\
\label{JMN-1metric} 
\end{eqnarray}
where the positive constant $M_0$ should always be less than 1, $R_b$ is the radius of the boundary of the JMN1 spacetime where it is smoothly matched with the external Schwarzschild spacetime. 

Now, one can derive the slowly rotating JMN1 naked singularity spacetime by using eqs. (\ref{general_form}) and (\ref{JMN-1metric}) as,
\begin{widetext}
\begin{eqnarray}\label{17}
ds^2=-(1-M_0)\left(\frac{r}{R_b}\right)^\frac{M_0}{(1- M_0)} dt^2+\frac{dr^2}{1-M_0}+r^2\left(d\theta^2+\sin^2\theta d\phi^2\right) \nonumber \\
-2a\sin^2\theta\left[\left(\frac{r}{R_b}\right)^{\frac{M_0}{2(1- M_0)}}-(1-M_0)\left(\frac{r}{R_b}\right)^{\frac{M_0}{(1- M_0)}}\right]dtd\phi.
\label{eq:rotating_JMN}
\end{eqnarray}
\end{widetext}
The above slowly rotating JMN1 spacetime can be matched with the exterior slowly rotating Kerr spacetime at $r=R_b$, 
\begin{eqnarray}
\label{eq:slowly_rotating_kerr}
ds^2 = &-&\left(1-\frac{M_0 R_b}{r}\right) dt^2+\frac{dr^2}{\left(1-\frac{M_0 R_b}{r}\right)}\nonumber\\&+&r^2\left(d\theta^2+\sin^2\theta d\phi^2\right)
-\frac{2aM_0R_b}{r}\sin^2\theta dt~d\phi\nonumber \\
\end{eqnarray}
where $M=\frac{M_0 R_b}{2}$ is the Schwarzschild mass of the slowly rotating compact object.

In order to describe the causal structure of JMN1 naked singularity spacetime, in Fig.~(\ref{fig1}), we show the Penrose diagram of the same.
In Figs.~\ref{re3} and \ref{re4}, we show Penrose diagrams of JMN1 interior spacetime for $ M_{0} = 1/3 $, $M_{0} = 3/4$ respectively. The pink dashed line in Figs.~\ref{re3} and \ref{re4}, shows the matching of internal JMN1 spacetime with external Schwarzschild spacetime at the matching radius $r= R_b$.
In those Penrose diagrams, the red dashed line shows the singularity of spacetime and the black line represents the past and future null infinities in Schwarzschild spacetime. Therefore, the area in between the pink dashed line and the black line represents Schwarzschild spacetime, and the remaining area in between the red dashed line and the pink dashed line represents JMN1 spacetime.

\subsection{Null naked singularity spacetime}
As we know, if the singularity is covered by trapped surfaces, then no information from the singularity can reach the faraway observer. Since this singularity is not causally connected to any other spacetime points, it is known as space-like singularity. There are other types of singularities that can be formed as the end state of the gravitational collapse of an inhomogeneous matter cloud, such gravitationally strong singularities are connected with other spacetime points
and these are known as nulllike and timelike singularities \cite{Joshi:2011zm,Joshi2020,Dey:2020bgo}. 

In \cite{Joshi2020}, we propose a new spherically symmetric and static nulllike naked singularity spacetime which asymptotically resembles the Schwarzschild spacetime. 
The line element of the nulllike naked singularity spacetime can be written as, 
\begin{equation}
ds^2 = -\frac{dt^2}{\left(1+\frac{M}{r}\right)^2}+\left(1+\frac{M}{r}\right)^2dr^2 + r^2d\Omega^2\,\, , 
\label{eq3}
\end{equation}
where $M$ is the ADM (Arnowitt, Deser \& Misner) mass and we show that the Kretschmann scalar and Ricci scalar blow up at the center $r=0$ \cite{Joshi2020}. One can easily verify that the strong, weak and null energy conditions are satisfied for the above spacetime. The Penrose diagram of the above nulllike naked singularity spacetime is shown in (Fig.~(\ref{fig2})). The main interesting thing is that if any null geodesic emanates from the past null infinity, the photon would be infinitely red-shifted with respect to the asymptotic observer. Now using the general expression of the slowly rotating metric (Eq.~\ref{general_form}), we can write the slowly rotating version of the above spacetime as,
\begin{eqnarray}\label{rotnull}
 ds^2 = -\frac{dt^2}{\left(1+\frac{M}{r}\right)^2}&+& \left(1+\frac{M}{r}\right)^2 dr^2+r^2\left(d\theta^2+\sin^2\theta d\phi^2\right)\nonumber\\
 &-&\frac{2a M\left(M+2r\right)}{\left(M+r\right)^2}\sin^2\theta dtd\phi.
\end{eqnarray}
Note that, since it is asymptotically flat spacetime, one need not match the above spacetime to the slowly rotating exterior Kerr spacetime.
 
In the next two sections, we investigate the nature of the Lense-Thirring precession of a test gyroscope and the orbital precession of the timelike bound orbit in the slowly rotating JMN1 (Eq.~(\ref{eq:rotating_JMN})) and nulllike nake singularity (Eq.~(\ref{rotnull})) spacetimes.

\section{Lense-Thirring effects in slowly rotating JMN1 and null naked singularity spacetimes}
\label{LTeffect}
In 1918, Josef Lense and Hans Thirring described spin precession of test gyroscope in the gravitational field of a spinning spherical body. As we discussed before, one can predict the causal structures of the spacetime geometries by investigating the Lense-Thirring precession effect. Therefore, the Lense-Thirring precession phenomenon is very much important for distinguishing black holes from naked singularities.

 Here, we consider a simple scenario where the gyroscopes are held fixed at every point of the space and we investigate how the precession frequency of the spin of test gyroscope changes with the change of spatial points. Since the gyroscope is at rest in a stationary spacetime, it moves along an integral curve ($\gamma(t)$) of a time like killing vector field $K$. The 4-velocity ($\bar{u}$) of the gyroscope can be written in terms of $\bar{K}$ as,
\begin{eqnarray}
\bar{u}=\frac{1}{\sqrt{-K^{\alpha}K_{\alpha}}} \bar{K}.
\end{eqnarray}  
The four velocity of timelike curve $\gamma(\tau)$ can be written as $u=\dot{\gamma}$, where $\tau$ is the proper time and $u$ is the tangent to the $\gamma(\tau)$ with the condition $u^{\mu}u_{\mu}=-1$. The component of Fermi derivative of any vector in the direction of $u$ can be written as,
\begin{eqnarray}\label{21}
u^{\alpha} F_{\alpha} X^{\beta}=u^{\alpha} \nabla_{\alpha} X^{\beta} - X^{\alpha}a_{\alpha}u^{\beta}+X^{\alpha}u_{\alpha}a^{\beta}\,\, ,
\end{eqnarray}
where $a^{\alpha}$ is the acceleration which can be written as $a^{\alpha}=u^{\beta}\nabla_{\beta}u^{\alpha}$. Now, if we consider an orthonormal frame $\{e_i\}$ ($i=1,2,3$) perpendicular to $e_0=u$, then along with the condition
$e^{\alpha}_\mu e_{{\nu}{\alpha}}=\eta_{\mu\nu}$,
where $\eta_{\mu\nu}=diag(-1,1,1,1)$, we can write the components of co-variant derivative of orthonormal basis in the direction of four velocity $u$ as,
\begin{eqnarray}\label{22}
u^{\alpha}\nabla_{\alpha}e^{\beta}_i=e^{\alpha}_i a_{\alpha} u^{\beta} + \omega^j_i e^{\beta}_j\,\, ,
\end{eqnarray}
where $\omega^j_i=u^{\alpha}\nabla_{\alpha}e^j_{\beta} e^{\beta}_i$. Therefore, using the above Eqs. (\ref{21}) and (\ref{22}), the component of Fermi derivative of orthonormal basis in the direction of four velocity $u$ can be written as,
\begin{eqnarray}\label{23}
u^{\alpha}F_{\alpha}e_{\beta}=\omega^{\alpha}_{\beta}e_{\alpha}.
\end{eqnarray}
Let S be the expectation value of the spin operator for a particle or the spin vector of a rigid rotator like gyroscope, then $u^{\alpha}F_{\alpha}S=0$, and after expanding this equation, we get the following expression of the change in spin S with respect to proper time $\tau$,
\begin{eqnarray}\label{24}
\frac{dS^{\alpha}}{d\tau}=\omega^{\alpha}_{\beta} S^{\beta}\,\, ,
\end{eqnarray}
which shows that spin vector S would precess with some frequency. The angular frequency ($\omega_{ij}$) can be defined as,
\begin{eqnarray}
\omega_{ij}=\epsilon_{ijk}\Omega^k\,\, ,
\end{eqnarray}
where $\Omega^k$ is the Lense-Thirring precession frequency and $\epsilon_{ijk}$ is the levi-civita tensor. In terms of timelike killing vector field $\bar{K}$, $\omega_{ij}$ can be written as,
\begin{eqnarray}
\omega_{ij}=\frac{1}{\sqrt{-K^{\alpha}K_{\alpha}}} e^{\beta}_j K^{\delta}{\nabla}_{\delta} e_{i{\beta}},
\end{eqnarray}
and therefore, $\bar{\Omega}$ can be expressed as,
\begin{eqnarray}
\bar{\Omega}=\frac{1}{2} \frac{1}{\sqrt{-K^{\alpha}K_{\alpha}}} *(\bar{K} \land d\bar{K})\,\, ,
\end{eqnarray}
where $*$ represents Hodge operator and ${\land}$ represents the exterior product. 
The above expression of the Lense-Thirring precession frequency for the spin vector S can be written as, 
\begin{eqnarray}\label{30}
\Omega_{\mu}=\frac{1}{2} \frac{1}{\sqrt{-K^{\alpha}K_{\alpha}}} \eta^{{\nu}{\rho}{\sigma}}_{\mu} K_{\nu} {\partial}_{\rho} K_{\sigma}\, ,
\end{eqnarray}
where ${\eta}$ is the canonical volume corresponding to the metric ($g_{{\mu}{\nu}}$). Since only for rotating spacetime $u^{\alpha}F_{\alpha}e_{\beta}\neq0$, the Lense-Thirring precession of a gyroscope is possible only in rotating spacetimes. For the stationary test gyroscope, the timelike killing vector ($\bar{K}$) mentioned above can be expressed as,
$\bar{K}=g_{tt}dt+g_{ti}dx^i$. Therefore, Eq.~(\ref{30}) becomes,
\begin{eqnarray}\label{31}
\Omega=\frac{g_{tt}}{2\sqrt{-g}} \epsilon_{ijk} \bigg(\frac{g_{ti}}{g_{tt}}\bigg)_{,j} \bigg(\partial_k-\frac{g_{kt}}{g_{tt}}\partial_t\bigg)\,\, ,
\end{eqnarray}
\begin{figure*}
\centering
\subfigure[Lense-Thirring precession in slowly rotating Kerr spacetime at $\theta=0\degree$.]
{\includegraphics[width=70mm]{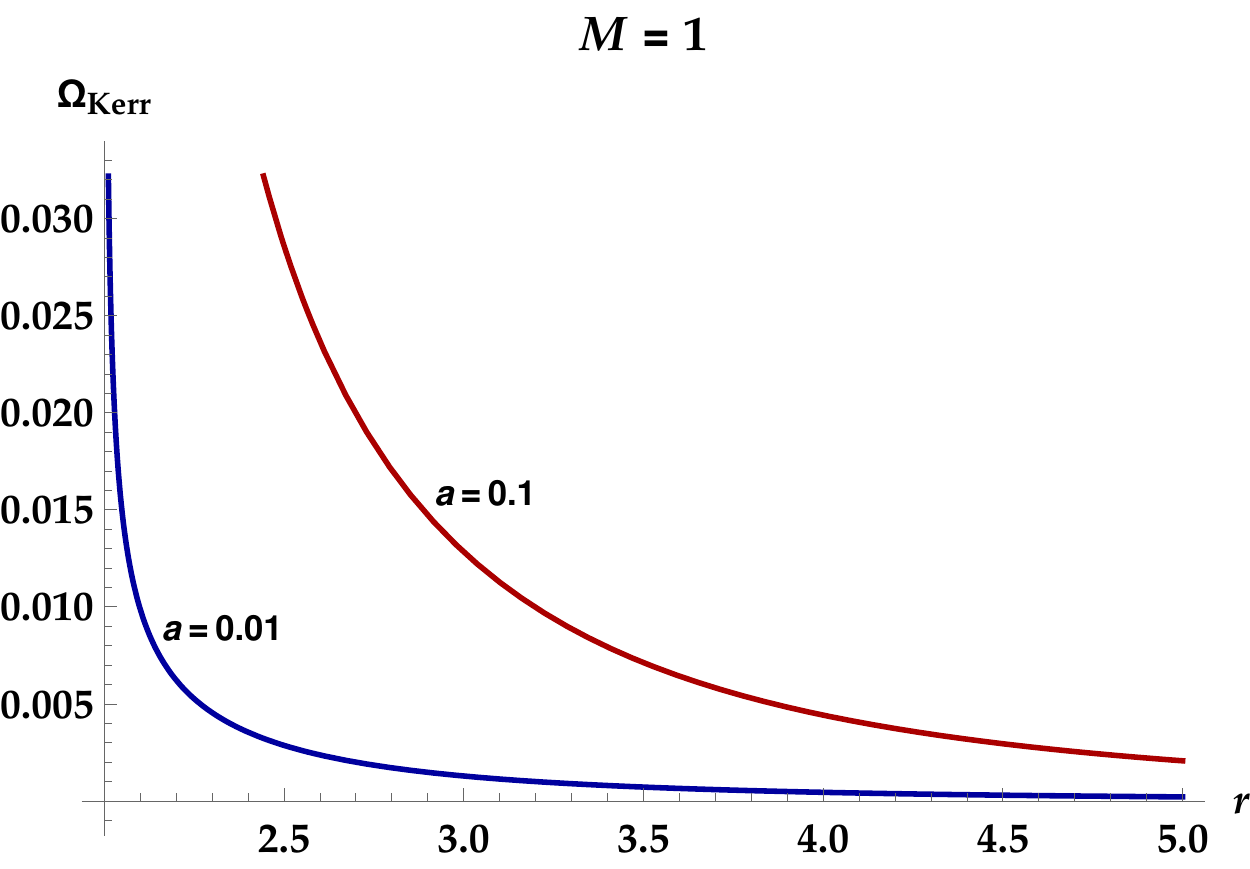}\label{re5}}
\subfigure[Lense-Thirring precession in slowly rotating Kerr spacetime at $\theta=90\degree$.]
{\includegraphics[width=70mm]{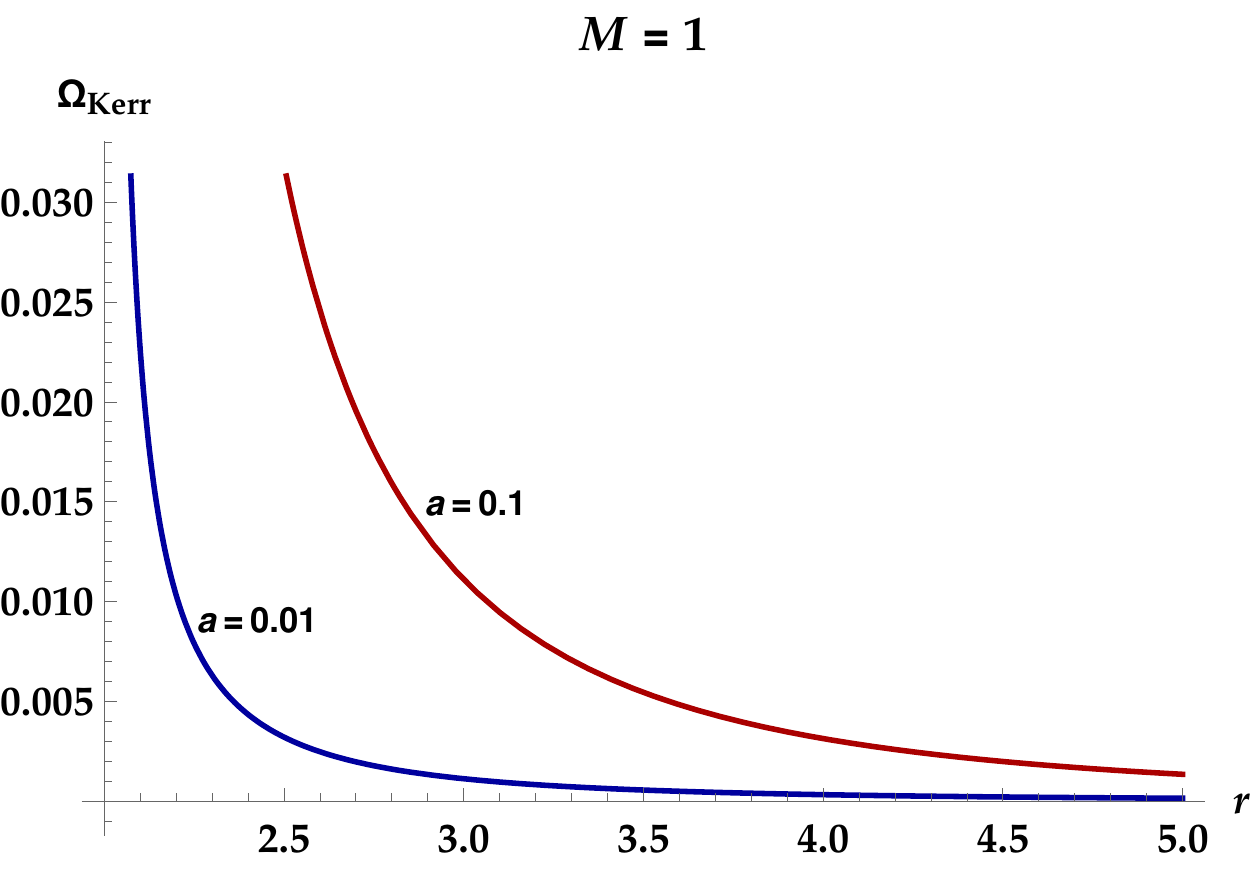}\label{re6}}
\subfigure[Lense-Thirring precession in slowly rotating JMN1 spacetime at $\theta=0\degree$.]
{\includegraphics[width=70mm]{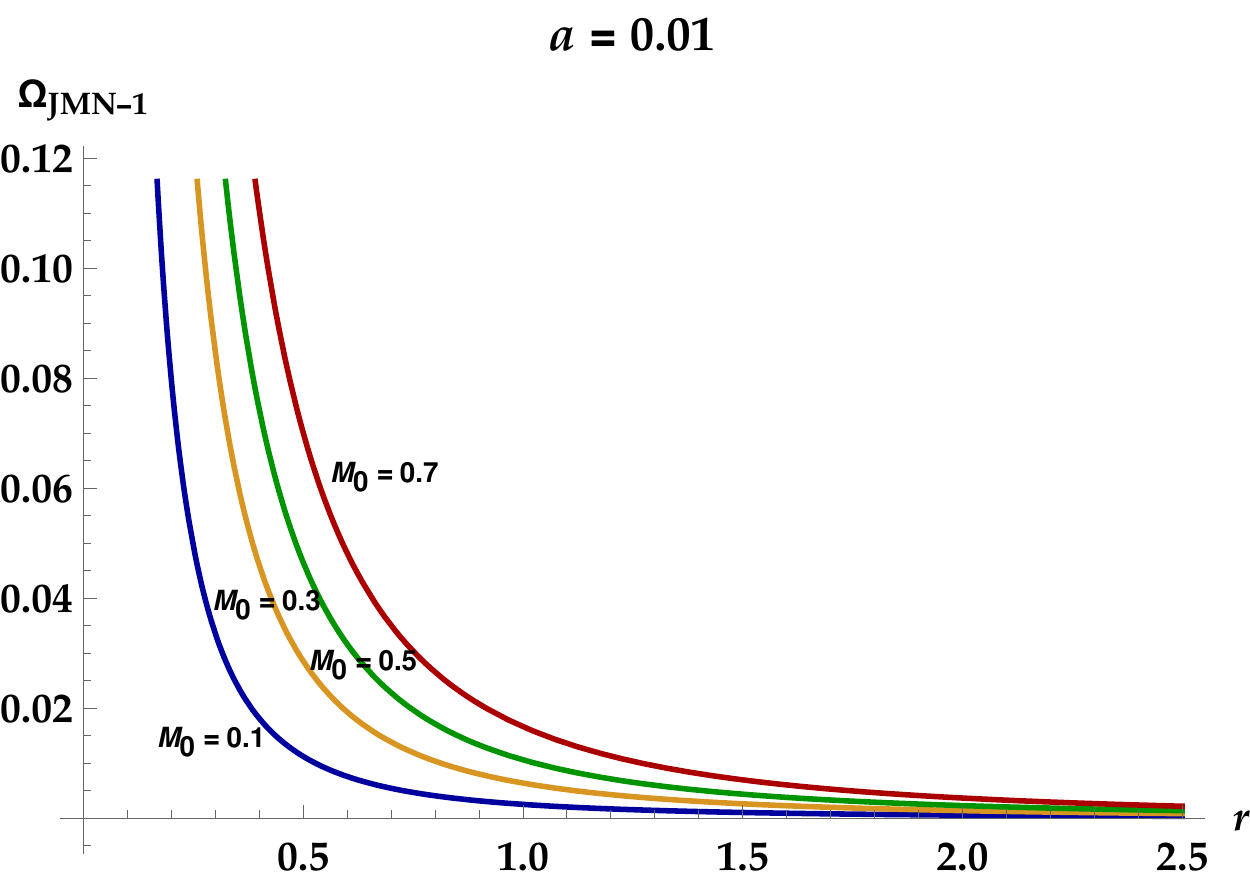}\label{re7}}
\subfigure[Lense-Thirring precession in slowly rotating JMN1 spacetime at $\theta=90\degree$.]
{\includegraphics[width=70mm]{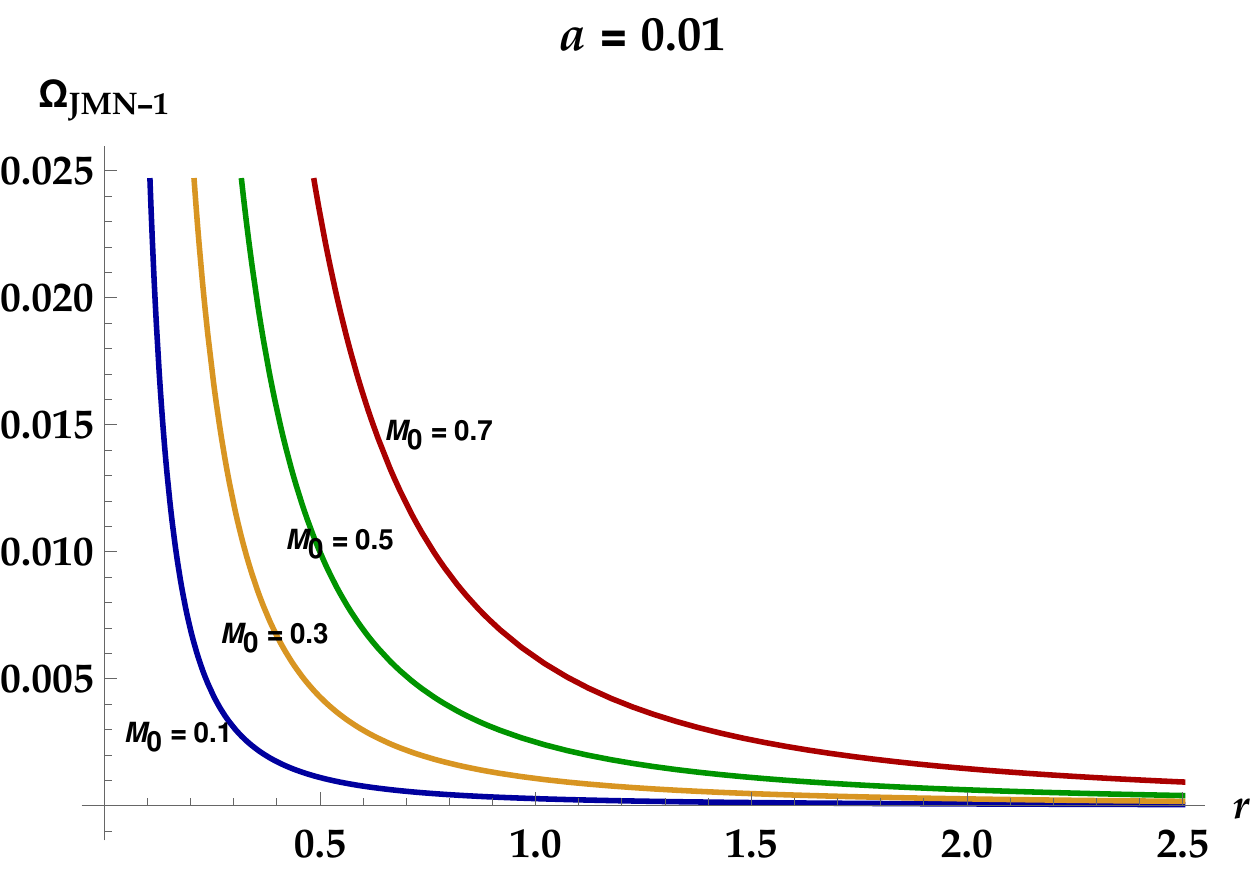}\label{re8}}
\subfigure[Lense-Thirring precession in slowly rotating JMN1 spacetime at $\theta=0\degree$.]
{\includegraphics[width=70mm]{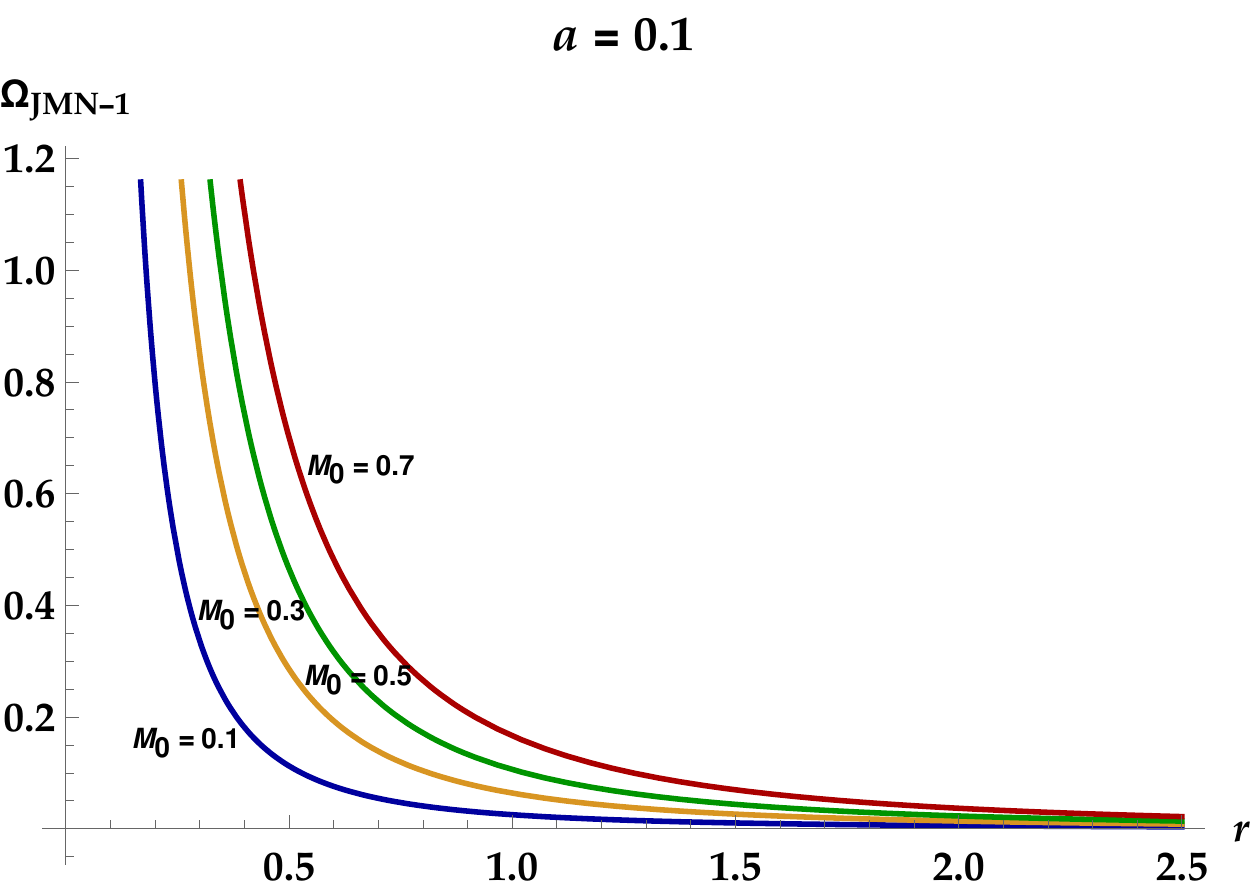}\label{re9}}
\subfigure[Lense-Thirring precession in slowly rotating JMN1 spacetime at $\theta=0\degree$.]
{\includegraphics[width=70mm]{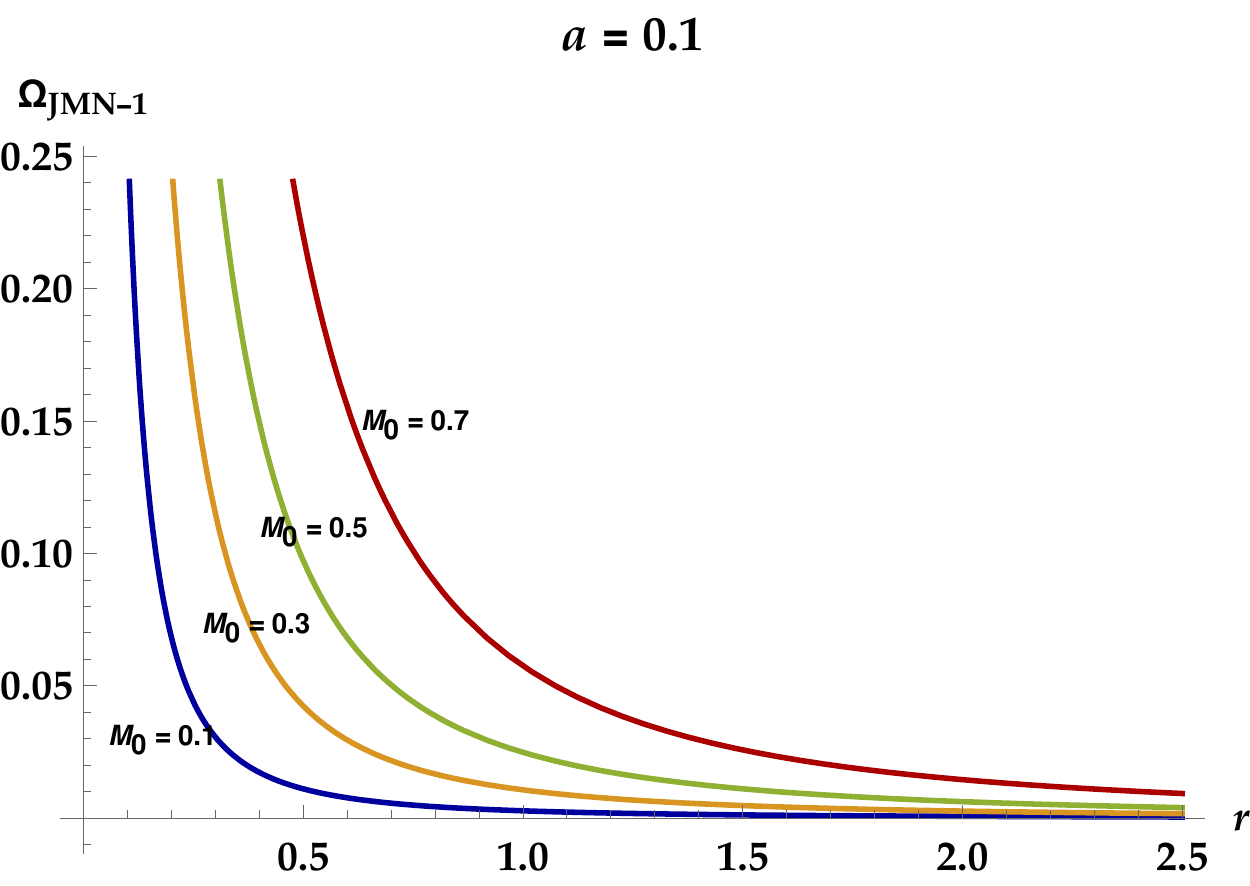}\label{re10}}
\subfigure[Lense-Thirring precession in slowly rotating null naked singularity spacetime at $\theta=0\degree$.]
{\includegraphics[width=70mm]{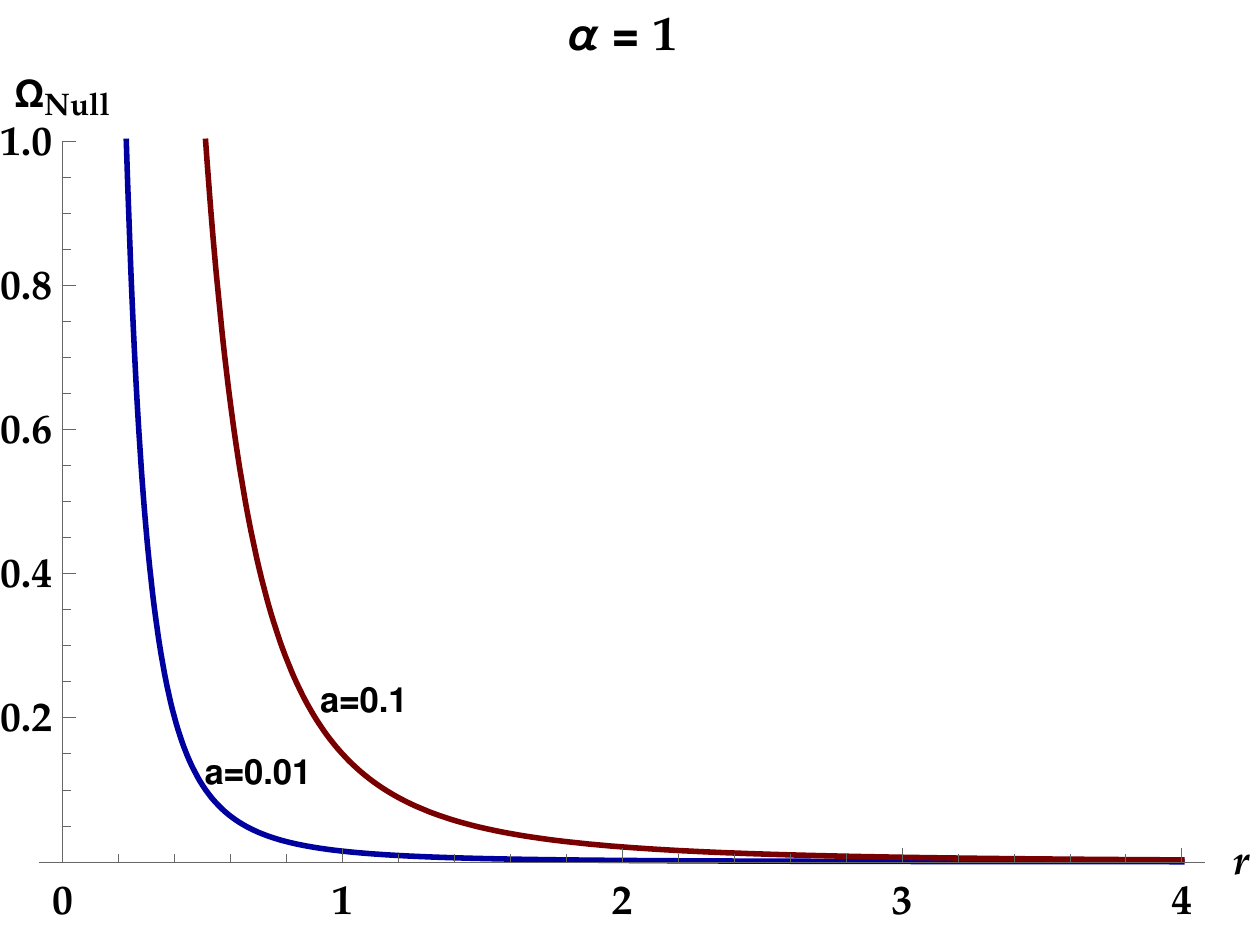}\label{re11}}
\subfigure[Lense-Thirring precession in slowly rotating null naked singularity spacetime at $\theta=90\degree$.]
{\includegraphics[width=70mm]{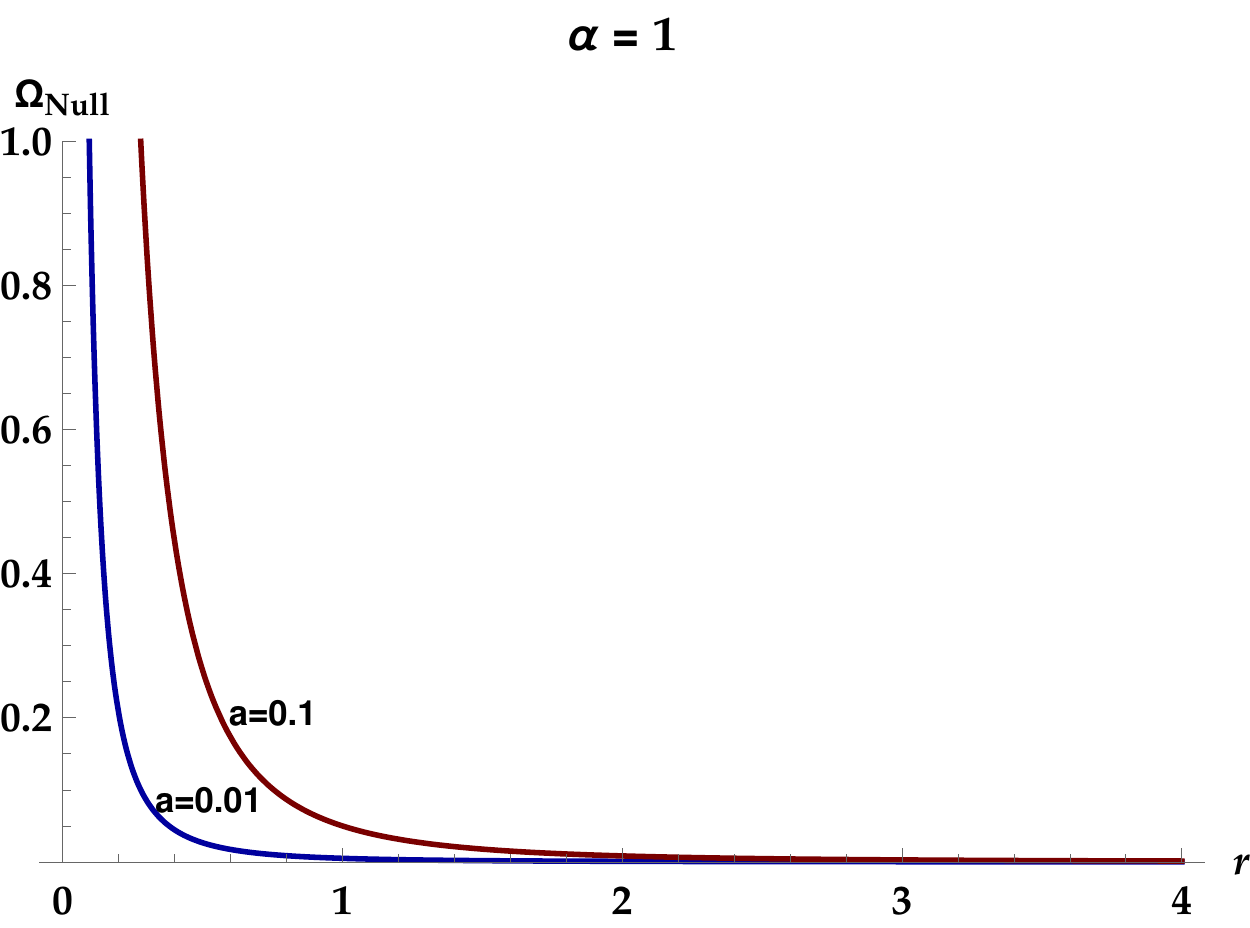}\label{re12}}
 \caption{Figure shows Lense-Thirring precession frequency ($\Omega$) in Kerr, JMN1 and null like naked singularity spacetimes for different values of spin parameter ($a$).}\label{fig3}
\end{figure*}
which also can be reduced to the following form
\begin{eqnarray}\label{32}
\Omega=\frac{g_{tt}}{2\sqrt{-g}} \bigg(\frac{g_{t{\phi}}}{g_{tt}}\bigg)_{,r} \partial_{\theta} - \frac{g_{tt}}{2\sqrt{-g}} \bigg(\frac{g_{t{\phi}}}{g_{tt}}\bigg)_{,{\theta}} \partial_r\,\, ,  
\end{eqnarray}
where we consider the general line element of a slowly rotating spacetime as
\begin{eqnarray}\label{genrotspt}
    ds^2=-g_{tt}(r)dt^2&+&g_{rr}(r)dr^2+g_{\theta\theta}(r)d\theta^2\nonumber\\&+&g_{\phi\phi}d\phi^2
    -2g_{t\phi}(r)dtd\phi\,\, .
\end{eqnarray}
The magnitude of the spin precession frequency can be written as, 
\begin{eqnarray}\label{34}
\Omega=\sqrt{g_{{\theta}{\theta}}(\Omega_{\theta})^2+g_{rr}(\Omega_r)^2}\,\, ,
\end{eqnarray}
where $\Omega_{\theta}=\frac{g_{tt}}{2\sqrt{-g}} \bigg(\frac{g_{t{\phi}}}{g_{tt}}\bigg)_{,r}$ and $\Omega_r = \frac{g_{tt}}{2\sqrt{-g}} \bigg(\frac{g_{t{\phi}}}{g_{tt}}\bigg)_{,{\theta}}$.

Now, from the above Eq.~({\ref{34}}), we can write the magnitude of the precession frequency for the slowly rotating Kerr spacetime (Eq.~(\ref{eq:slowly_rotating_kerr})) in terms of the spin parameter $a$, mass $M$, radial distance $r$ and the angle ${\theta}$ as,
\begin{eqnarray}\label{35}
\Omega_{Kerr}=\sqrt{\frac{a^2M^2((8M-4r)\cos^2\theta-r\sin^2\theta)}{(2M-r)r^2(r^3(-2M+r)+4a^2M^2\sin^2\theta)}},\nonumber\\
\end{eqnarray}
where the above expression is only valid for the radial distance greater than the radius of the ergosphere.
In Fig.~(\ref{fig3}), we show Lense-Thirring precession frequency ($\Omega$) in slowly rotating Kerr, JMN1, and null naked singularity spacetimes. In \cite{Chakraborty:2016ipk}, the Lense-Thirring effect in Kerr spacetime is elaborately discussed.  In this paper, we first reproduce the results of the spin precession of a test gyroscope in Kerr spacetime, and then, we compare the same with that for JMN1 and nulllike naked singularity spacetimes. We get novel features of the spin precession frequency of a test gyroscope in the naked singularity spacetimes and that can be useful to distinguish them from a black hole. Figs.~(\ref{re15},\ref{re16}) show the behaviour of Lense-Thirring precession in Kerr spacetime for $\theta=0 $ and $\theta = \frac{\pi}{2}$ respectively and we can see that the Lense-Thirring precession frequency in slowly rotating Kerr spacetime monotonically increases as radial distance decreases and becomes infinity at $r=2M$, where we consider $M=1$.  
On the other hand, the expression of the Lense-Thirring precession frequency for the Null and  JMN1 naked singularities are,
\begin{eqnarray}\label{37}
\Omega_{Null}=\sqrt{{\frac{a^2 {M}^2 ((2r+{M})^2\cos^2{\theta}+r^2 \sin^2{\theta})}{r^2((r+{M})^2 r^4+a^2M^2(M+2r)^2)\sin^2\theta)}}}\,\, ,\nonumber\\
\end{eqnarray}
\begin{widetext}
\begin{eqnarray}\label{36}
\Omega_{JMN-1}=\frac{1}{4} \sqrt{\frac{a^2 \left(16(-1+M_0)\left(1-\left(\frac{r}{R_b}\right)^{\frac{M_0}{2-2M_0}}+M_0\left(\frac{r}{R_b}\right)^{\frac{M_0}{2-2M_0}}\right)^2\cos^2\theta-M_0^2\sin^2\theta\right)}{(-1+M_0)r^2\left(-((-1+M_0)r^2)+a^2\left(1-\left(\frac{r}{R_b}\right)^{\frac{M_0}{2-2M_0}}+M_0\left(\frac{r}{R_b}\right)^{\frac{M_0}{2-2M_0}}\right)^2\sin^2\theta\right)}}.
\end{eqnarray}
\end{widetext}
\begin{figure*}
\centering
\subfigure[Lense-Thirring precession in slowly rotating Kerr spacetime.]
{\includegraphics[width=70mm]{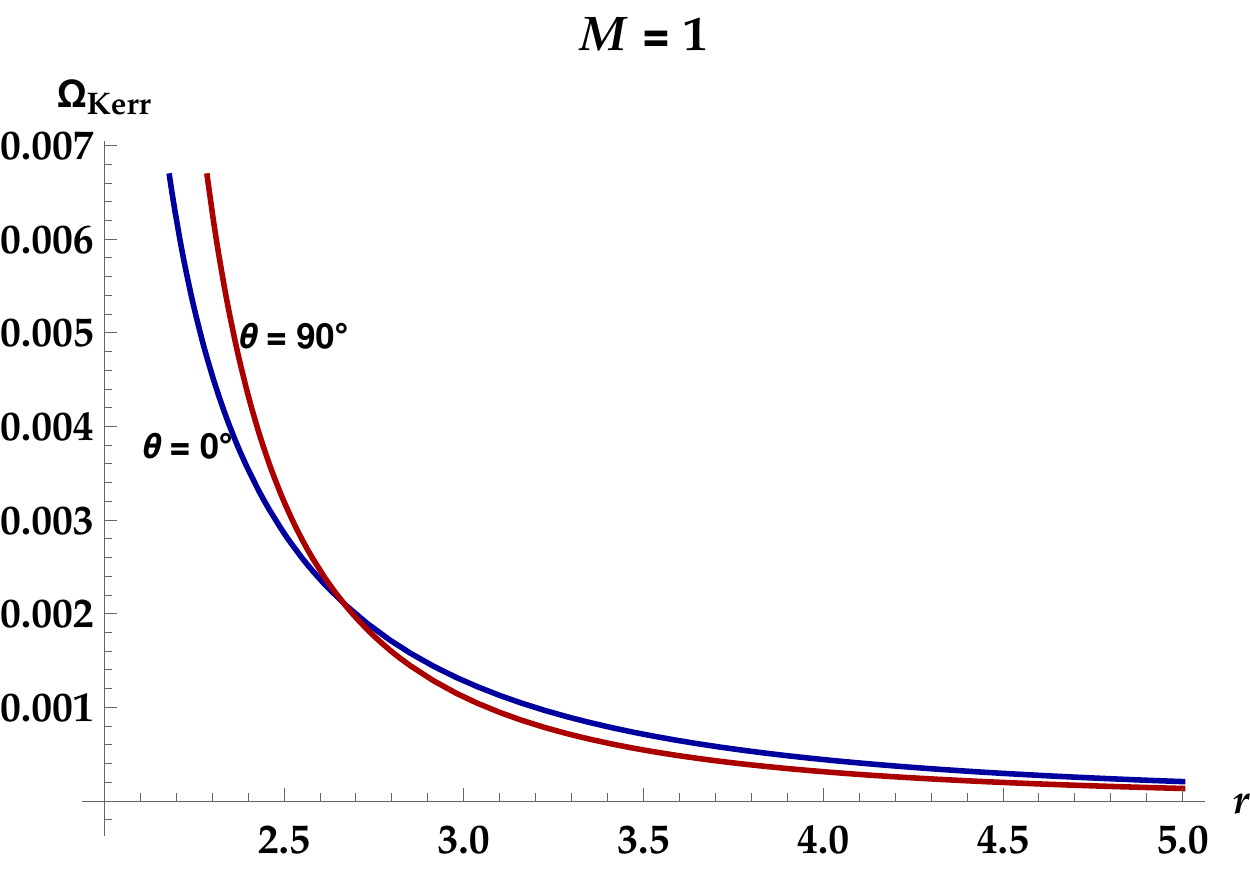}\label{re66}}
\subfigure[Lense-Thirring precession in slowly rotating JMN1 spacetime.]
{\includegraphics[width=70mm]{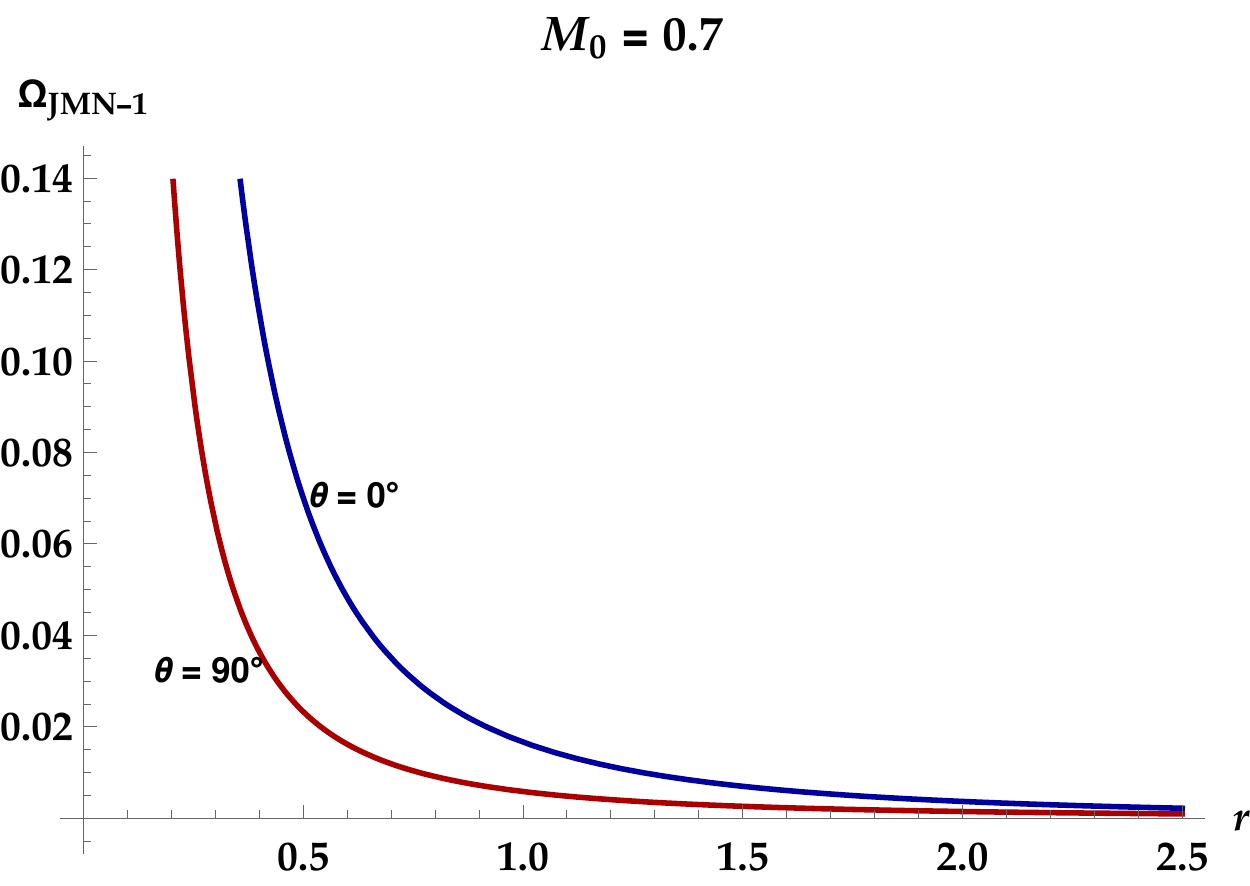}\label{re77}}
\subfigure[Lense-Thirring precession in slowly rotating JMN1 spacetime.]
{\includegraphics[width=70mm]{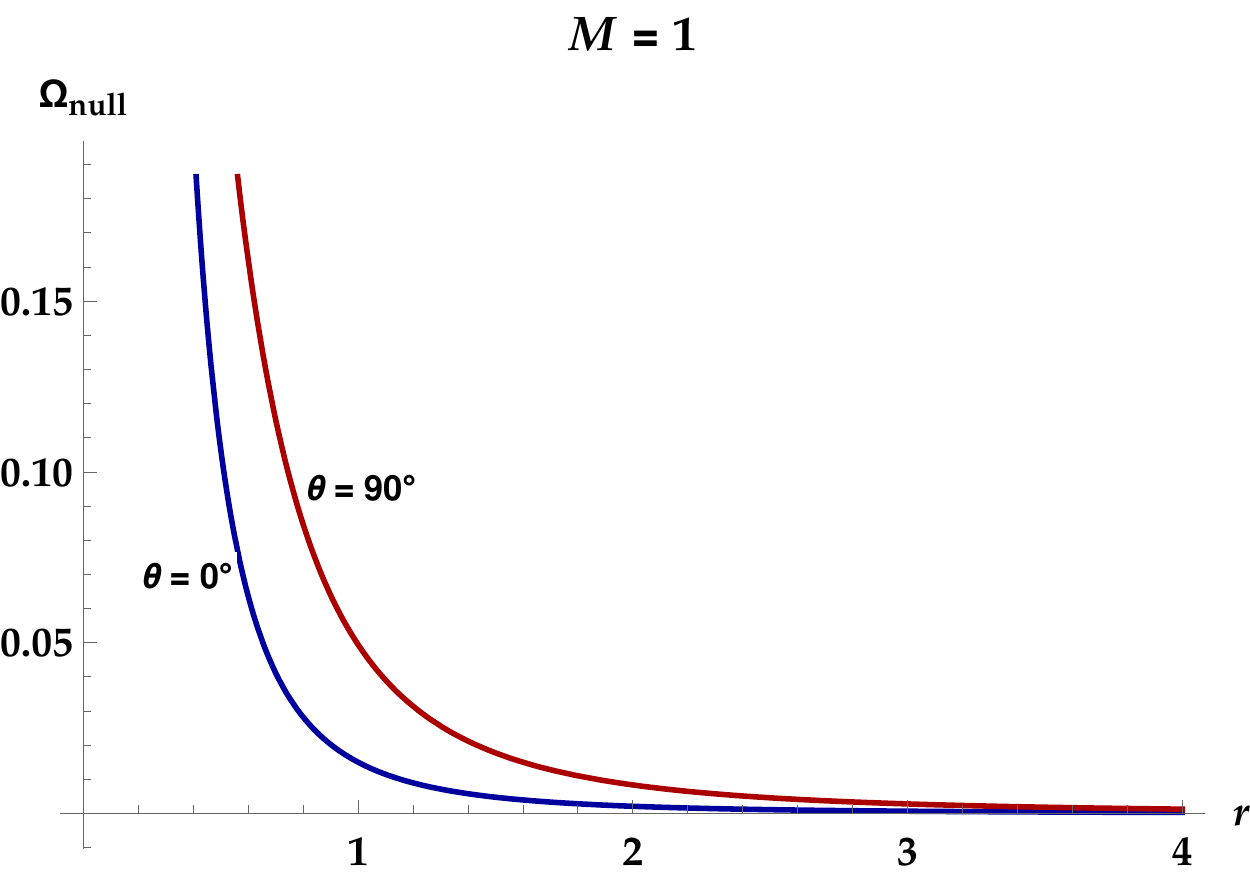}\label{re88}}
 \caption{Figure compares Lense-Thirring precession frequency ($\Omega$) at poles with that at the equatorial plane for Kerr, JMN1 and null like naked singularity spacetimes with fixed spin parameter $a=0.01$.}\label{fig33}
\end{figure*}
Figs.~(\ref{re7}) and (\ref{re9}) show the behaviour of the spin precession frequency ($\Omega_{JMN1}$) at the pole (i.e., $\theta =0$) in slowly rotating JMN1 spacetime, where spin parameter $a=0.01$ and $a=0.1$ respectively. On the other hand, Figs.~(\ref{re8}) and (\ref{re10}) depicts the nature of the Lense-Thirring precession  for same values of spin parameters of the slowly rotating JMN1 spacetime at the equitorial plane (i.e., $\theta =\frac{\pi}{2}$). Unlike Kerr spacetime, in JMN1 spacetime, the precession frequency becomes infinity at the central singularity. In Figs.~(\ref{re11}, \ref{re12}), similar nature of the Lense-Thirring effect can be seen in null naked singularity spacetime. The spin precession frequency ($\Omega_{Null}$) in Null naked singularity spacetime also increases monotonically as radial distance decreases and blows up at the center. Fig.~(\ref{fig33}) depicts how $\Omega_{Kerr} (Fig.~(\ref{re66})), \Omega_{JMN1} (Fig.~(\ref{re77})), \Omega_{Null} (Fig.~(\ref{re88}))$ behave at pole ($\theta = 0$) and at equatorial plane ($\theta = \frac{\pi}{2}$), where we consider $a=0.01$. In Fig.~(\ref{re66}), it can be seen that in Kerr spacetime, near $r=2M$, $\Omega_{Kerr}$ at equatorial plane is greater than that at pole. However, after a certain radius, we can see an opposite behaviour of $\Omega_{Kerr}$. On the other hand, in JMN1 spacetime, $\Omega_{JMN1}$ at pole is always greater than that at equatorial plane ((Fig.~(\ref{re77}))), whereas in null naked singularity spacetime, $\forall r,~ \Omega_{Null}|_{\theta=\frac{\pi}{2}}>\Omega_{Null}|_{\theta=0}$ (Fig.~(\ref{re88})).

As we discussed above, in slowly rotating Kerr spacetime, $\Omega_{Kerr}$ blows up near a finite radius $r=2M$. However, $\Omega_{JMN1}$ and $\Omega_{null}$ blow up only at the central singularity. This distinguishable feature of the Lense-Thirring precession in the naked singularity spacetimes can be very useful observationally in order to distinguish a black hole from a naked singularity.  
\begin{figure*}
\centering
\subfigure[$M_0=0.3$, $R_b=6.66$,$E=-0.18$ and $a=0.01$.]
{\includegraphics[width=70mm]{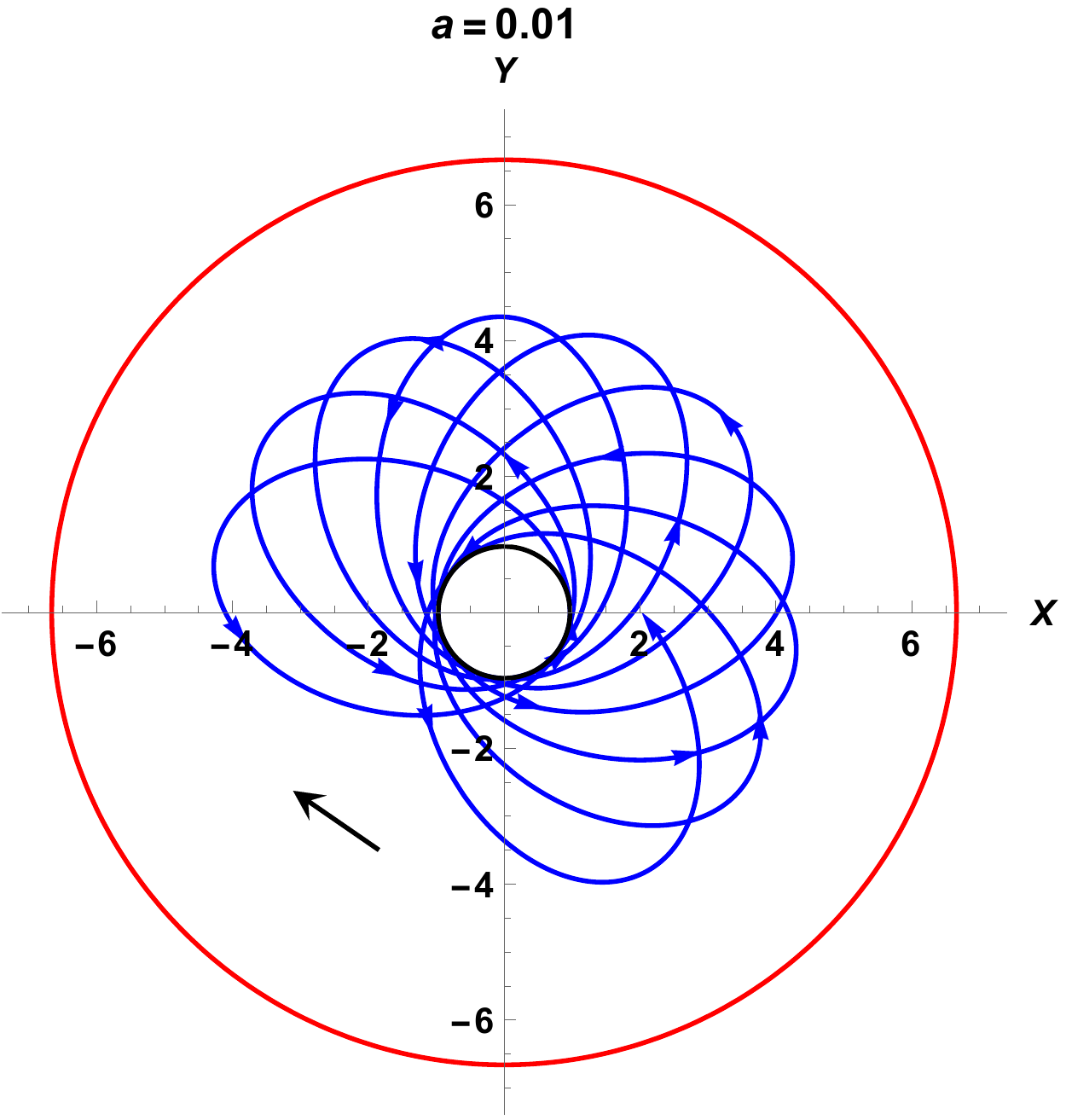}\label{re13}}
\subfigure[$M_0=0.4$, $R_b=5$,$E=-0.21$ and $a=0.01$.]
{\includegraphics[width=70mm]{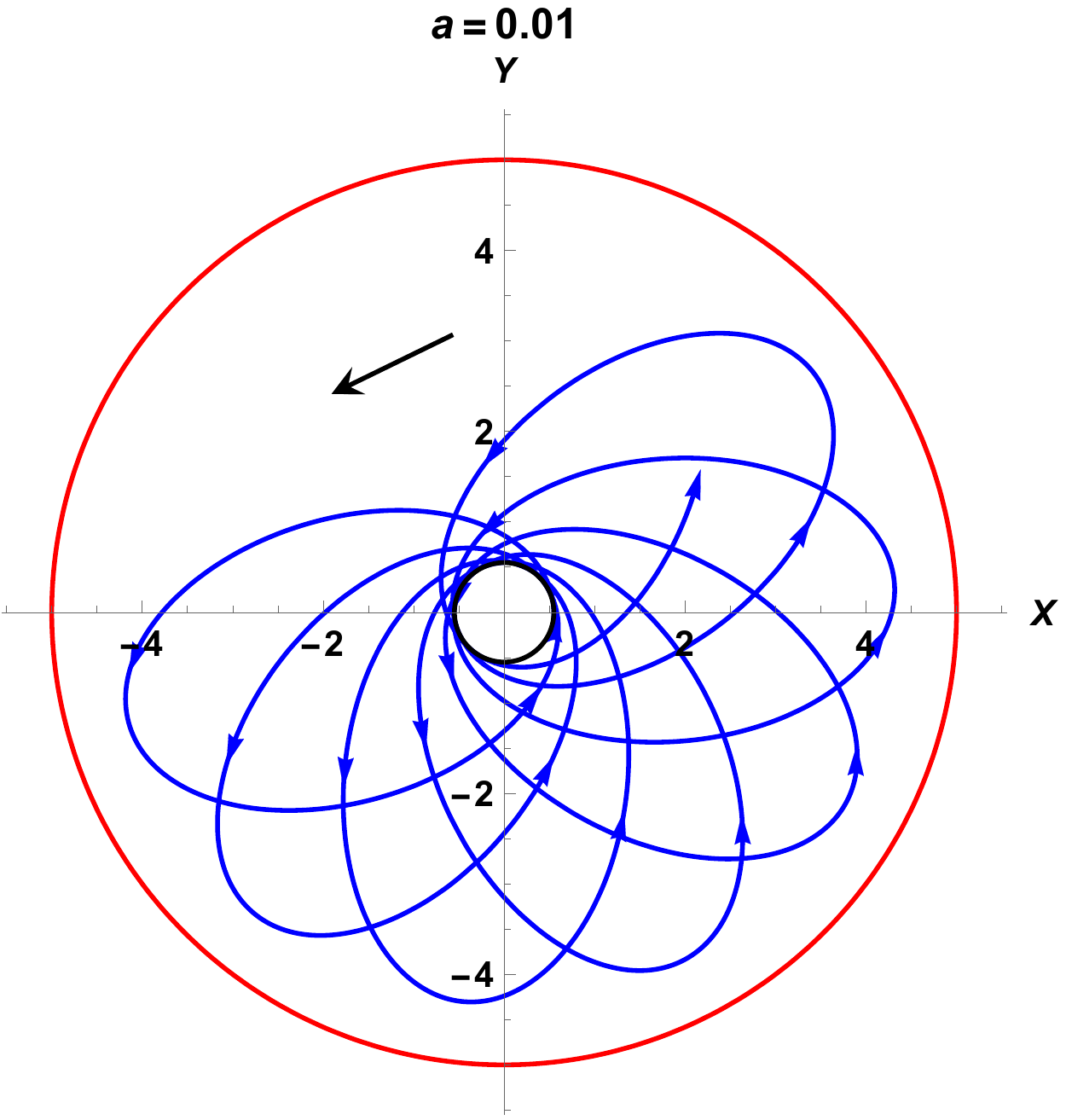}\label{re14}}
\subfigure[$M_0=0.3$, $R_b=6.66$,$E=-0.18$ and $a=0.1$.]
{\includegraphics[width=70mm]{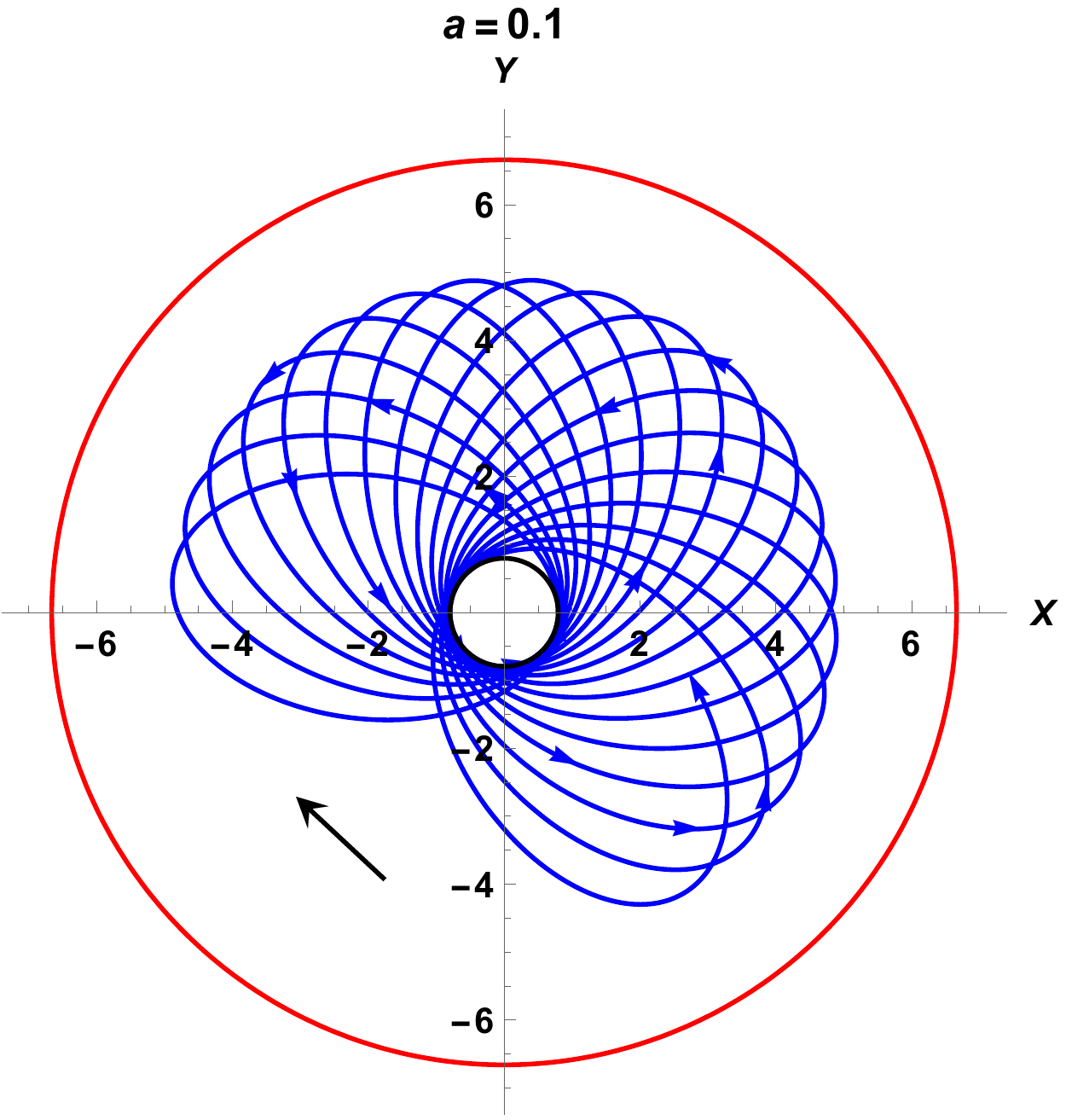}\label{re15}}
\subfigure[$M_0=0.4$, $R_b=5$,$E=-0.21$ and $a=0.1$.]
{\includegraphics[width=70mm]{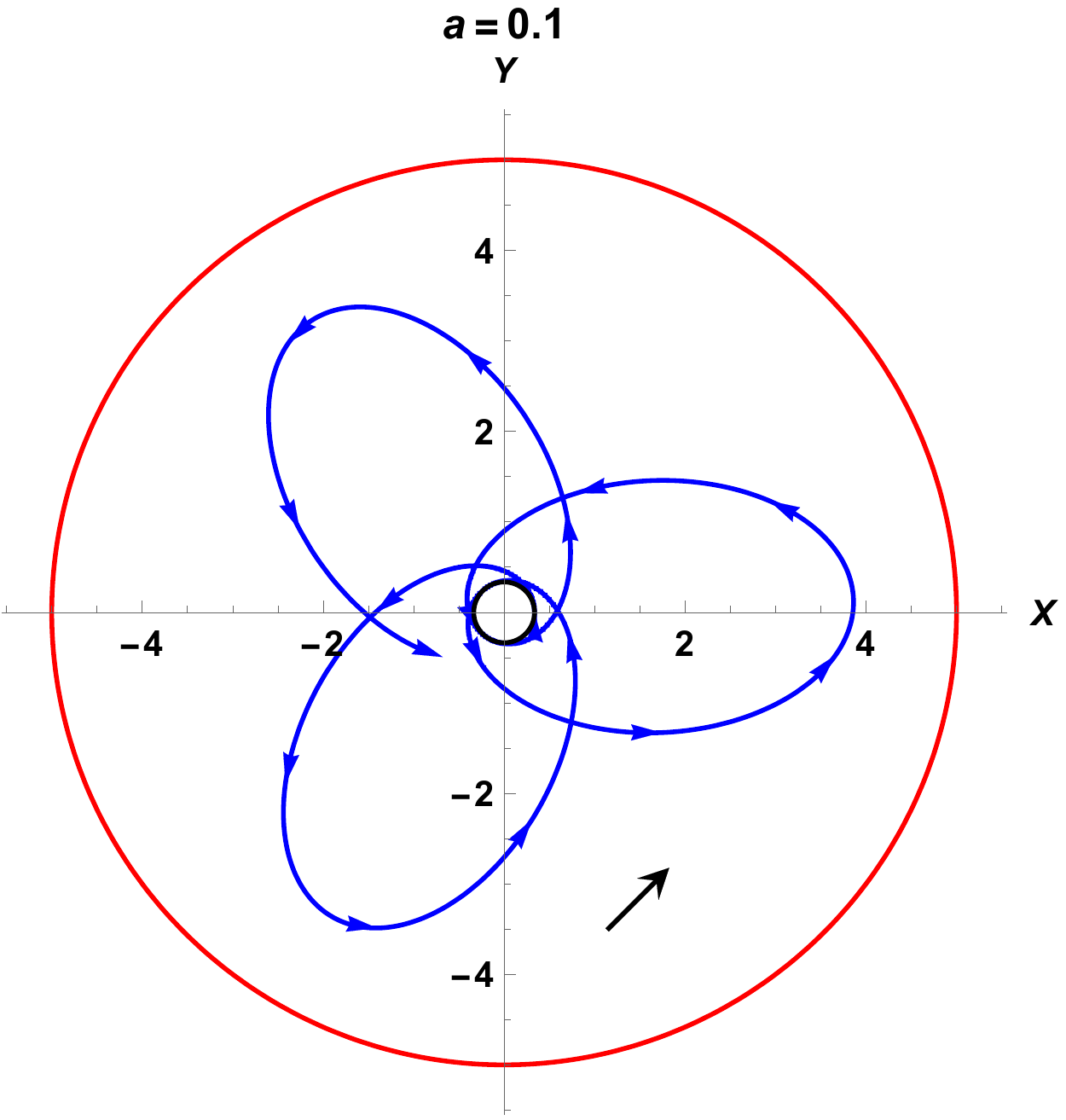}\label{re16}}
\caption{Relativistic orbits of a test particle in slowly rotating JMN1 spacetime for $L=1$, $M=1$. Where, black circle denotes for the minimum approach of a particle and the red circle represents the boundary of the JMN1 spacetime $R_b$.}\label{jmn1orbit}
\end{figure*}
\begin{figure*}
\subfigure[$E=-0.013$ and $a=0.01$.]
{\includegraphics[width=70mm]{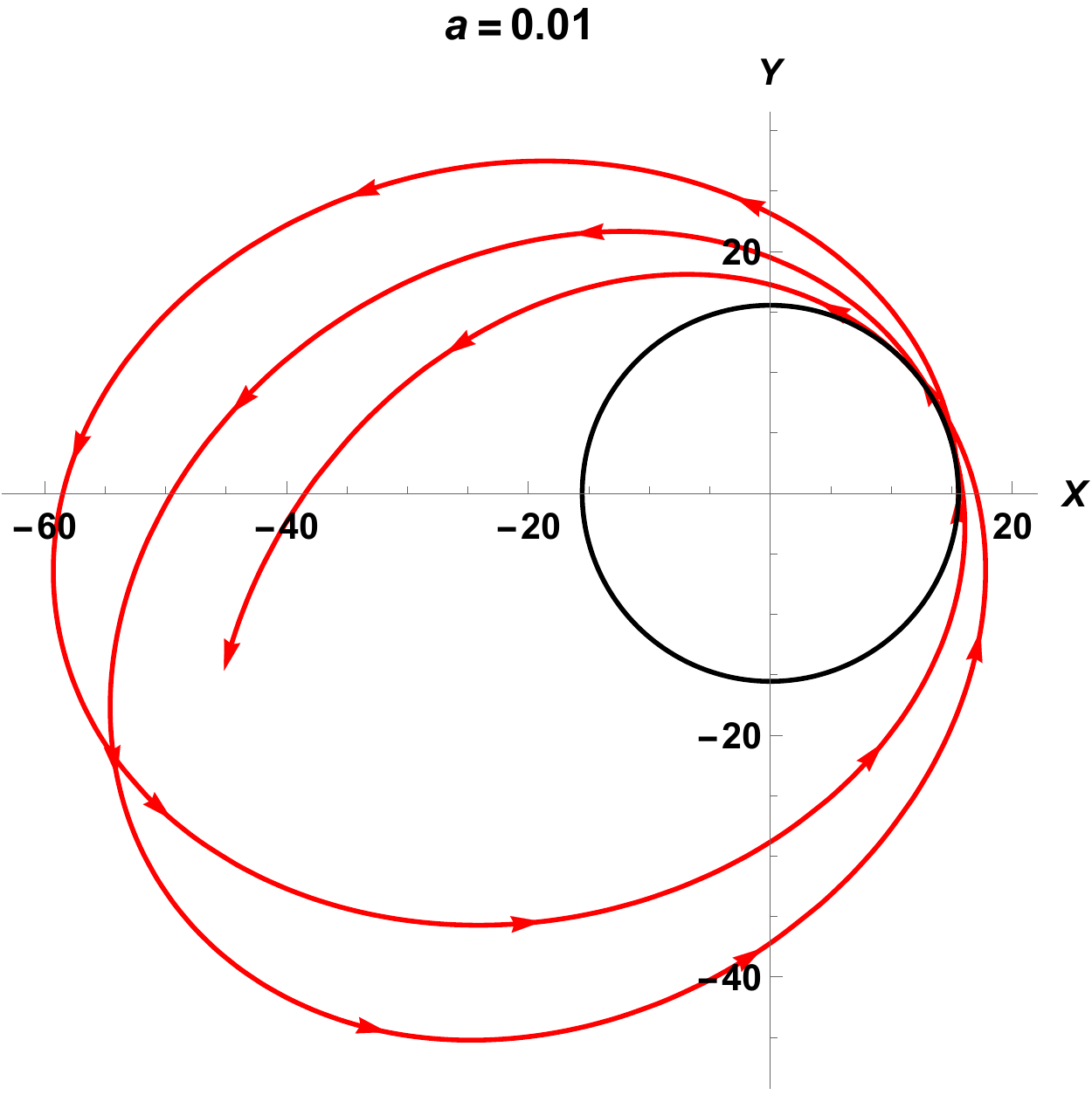}\label{re17}}
\subfigure[$E=-0.013$ and $a=0.1$.]
{\includegraphics[width=70mm]{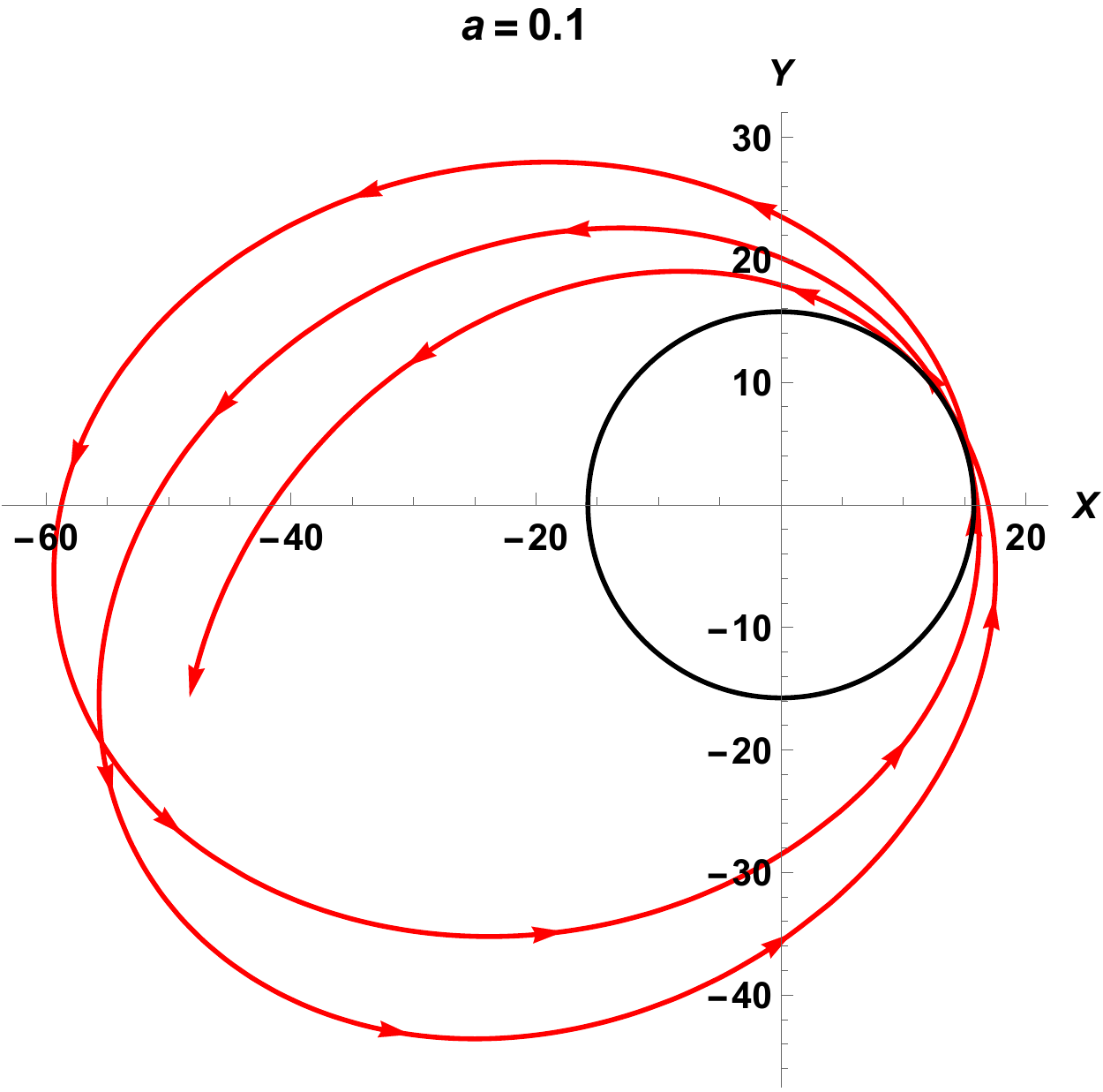}\label{re18}}
\caption{Relativistic orbits of a test particle in slowly rotating null naked singularity spacetime for $L=5$, $M=1$. Where, black circle denotes for minimum approach of a particle. }\label{nullorbit}
\end{figure*}
\section{Timelike geodesics in slowly rotating spacetimes}
\label{orbit1}
In this section, we analyse the timelike geodesics of the bound orbits in the slowly rotating black hole and naked singularity spacetimes. First, we review the general procedure for deriving an orbit equation. As it is mentioned in the previous section, the general form of the stationary, axially symmetric, and slowly rotating spacetime can be written as,
\begin{eqnarray}
    ds^2=-g_{tt}(r)dt^2&+&g_{rr}(r)dr^2+g_{\theta\theta}(r)d\theta^2\nonumber\\&+&g_{\phi\phi}d\phi^2
    -2g_{t\phi}(r)dtd\phi\,\, .
\end{eqnarray}
The above general spacetime is independent of $t$ and $\phi$, therefore, the energy $e$ and the angular momentum $L$ (about the symmetry axis) per unit rest mass are conserved. Using these symmetry properties and the four-velocity normalization condition for the timelike geodesics ($u^{\alpha}u_{\alpha}=-1$), we can derive the general expression of the total relativistic energy as,  
\begin{eqnarray}
E = \frac{1}{2} (e^2 - 1) = \frac{1}{2} (u^r)^2 + V_{eff}(r)\,\, ,\label{E1}
\end{eqnarray}
where $u^r$ is radial component of the four velocity $u^{\mu}$, and the effective potential can be described as,
\begin{eqnarray}
V_{eff}(r)=\frac{1}{2 g_{rr}} + \frac{e^2-1}{2} + \frac{L^2 g_{tt}+2eL g_{t\phi}-e^2 g_{\phi\phi}}{2 g_{rr}(g_{t\phi}^2+g_{tt}g_{\phi\phi})}
\label{Veff general}
\end{eqnarray}
Note that, for simplicity, we have considered the equatorial plane ($\theta=\frac{\pi}{2}$) for the analysis of the timelike bound orbits. Using the components of the metric tensor $g_{\mu\nu}$ given in Eqs. (\ref{eq:rotating_JMN}) and (\ref{general_form}), respectively the effective potentials for the JMN1 and null naked singularity spacetimes can be written by as,
\begin{widetext}
\begin{eqnarray}
V_{eff}|_{JMN1}=\frac{1}{2}\left[e^2 - M_0 + \frac{(-1 + M_0)\left(P(e,a,r,L)\right)\left(\frac{r}{R_b}\right)^{\frac{M_0}{-1+M_0}}}{-(-1+M_0)r^2+\left(a+a(-1+M_0)\left(\frac{r}{R_b}\right)^{\frac{M_0}{(2-2M_0)}}\right)^2} \right]\,\, ,
\end{eqnarray}
where $P(e,a,r,L)=e^2 r^2 + 2 a e L \left(\frac{r}{R_b}\right)^{\frac{M_0}{(2-2M_0)}} + L (2 a e + L)(-1+M_0)\left(\frac{r}{R_b}\right)^{\frac{M_0}{1-M_0}}$ and
\begin{eqnarray}
V_{eff}|_{null}=\frac{1}{2}\left[-1+\frac{r^2}{(M+r)^2}+\frac{(Lr^2+2aeM(M+2r))^2}{r^4(M+r)^2+4a^2M^2(M+2r)^2}\right]\,\, .
\end{eqnarray}
\end{widetext}
For the timelike bound circular orbits, the following condition is always satisfied,
\begin{eqnarray}
 \frac{dV_{eff}}{dr}|_{r_b} = 0 ;        \frac{d^2 V_{eff}}{dr^2}|_{r_b}>0\,\, ,
    \label{con1}
\end{eqnarray}
where at $r=r_b$ the effective potential has a minimum.
Therefore, for the timelike bound orbits, the total energy ($E$) of the particle is greater than or equal to the minimum of the effective potential $V_{min}\leq E$. By using the bound timelike orbit conditions, we can determine the shape and nature of the orbits from the following differential equation of orbit, 
\begin{widetext}
\begin{eqnarray}
\bigg(\frac{dr}{d\phi}\bigg)^2=-\frac{(g_{t\phi})^2+g_{tt}g_{\phi\phi}}{g_{rr}(e g_{t\phi} + L g_{tt})^2} [(g_{t\phi})^2-e^2 g_{\phi\phi}+2eL g_{t\phi}+g_{tt}(L^2+g_{\phi\phi})].
\label{general orbit}
\end{eqnarray}
\end{widetext}
Now, we can derive the form of an orbit equation by differentiating the above equation with respect to $\phi$ \cite{Bam2020,Bambhaniya:2021ybs}.
One can get the orbit equations for the slowly rotating JMN1 and null naked singularity spacetimes given in Eqs.~ (\ref{eq:rotating_JMN}) and (\ref{general_form}). Since these are non-linear second-order differential equations, it is very difficult to solve analytically. Therefore, we solve it numerically and we depict the timelike bound orbits in Fig.~(\ref{jmn1orbit}) and Fig.~(\ref{nullorbit}).

In Fig.~(\ref{jmn1orbit}), we show the relativistic orbits of a test particle in slowly rotating JMN1 spacetime, where we consider $M_0=0.3$ and $M_0=0.4$ along with the different parameters' values (i.e., $a, E, R_b$). For all the cases the Schwarzschild mass $M=1$. 
One can see from Figs.~(\ref{re13}) and (\ref{re15}), the nature and shape of the orbits are different compare to the orbits shown in Figs.~(\ref{re14}) and (\ref{re16}). We find the timelike bound orbits of a particle in slowly rotating JMN1 spacetime can have negative as well as positive precession. As we discussed previously, for positive precession, orbit precesses in the direction of the particle motion, whereas for the negative precession, it precesses in the opposite direction of the particle motion. Figs.~(\ref{re13}, \ref{re15}) and Figs.~(\ref{re14}, \ref{re16}) respectively show the negative and positive precession of bound timelike orbits in slowly rotating JMN1 spacetime. In \cite{Bam2020}, it is shown that in Kerr spacetime, the negative precession of timelike bound orbits is not allowed. Therefore, the nature of the precession of the timelike bound orbits may be useful for distinguishing a rotating black hole from a rotating naked singularity.
Moreover, we find that the direction of the orbital shift or precession depends upon the values of $M_0$ and spin parameter $a$.
In \cite{Bambhaniya:2019pbr}, we derive the analytical expression of the orbital shift by using first-order eccentricity approximation for the spherically symmetric and static JMN1 spacetime. In that paper, we show that the negative precession occurs for $M_0<1/3$, while for $M_0>1/3$ we get positive precession. 
Now, if we introduce rotation by applying the Janis-Newman procedure in the static JMN1 spacetime, the range of $M_0$ for which we get negative or positive precession, would depend upon the value of the spin parameter $a$.
In Fig.~(\ref{nullorbit}), we show the relativistic orbits of a test particle in slowly rotating null naked singularity spacetime for $a=0.01$ (Fig.~(\ref{re17})) and $a=0.1$ (Fig.~(\ref{re18})), where we consider $M=1$, $L=5$, $E=-0.013$. From Figs.~(\ref{re17}) and (\ref{re18}), one can see that the positive precession occurs for the small values of the spin parameter.  
\section{Discussion and Conclusion}
Followings are the important results what we get in this paper,
\begin{itemize}
\item In this paper, we first use the Newman-Janis algorithm to get the rotating counter-parts of JMN1 and null naked singularity spacetimes. We show that the algorithm is not suitable to get the rotating version of a general spherically symmetric spacetime having any arbitrary value of the spin parameter. One has to use slow rotation approximation to get rid of this problem. Therefore, in this paper, we derive the metric of slowly rotating JMN1 and null naked singularity spacetime using Newman- Janis algorithm.
\item We investigate the nature of the spin precession of a stationary test gyroscope in the slowly rotating JMN1 and null naked singularity spacetime. We show that in both the naked singularity spacetimes, the Lense-Thirring precession frequency ($\Omega_{JMN1}, \Omega_{Null}$) monotonically increases as the radial distance from the central singularity decreases and it becomes infinity at the center. However, unlike slowly rotating naked singularity spacetimes, in Kerr spacetime, the spin precession frequency ($\Omega_{Kerr}$) blows up near the finite radial distance $r= 2M$.
\item In the Sec.~(\ref{orbit1}), we investigate the precession of the time like bound orbits in slowly rotating JMN1 and null naked singularity spacetimes. The novel feature of the timelike bound orbits in the rotating JMN1 spacetime is that it can precesses in the opposite direction of the particle motion (i.e., negative precession), which is not allowed in Kerr spacetime. On the other hand, in slowly rotating null naked singularity spacetime, only the positive precession is allowed. 
\end{itemize}
The Lense-Thirring precession and the orbital precession both are important physical phenomenon in the strong gravity region. Therefore, this two phenomena can be used to understand the spacetime geometry near an ultra high-density compact object. Hence, the distinguishable behaviour of the two types of precession in naked singularity spacetimes may be useful to differentiate a black hole from a naked singularity observationally.
\label{concludesec}

\end{document}